\documentclass{jpsj2}

\newcommand{\Bell}{\mbox{\boldmath $\ell \:$}}

\newcommand{\Bk}{\mbox{\boldmath $k \:$}}
\newcommand{\BR}{\mbox{\boldmath $R \:$}}

\newcommand{\Bi}{\mbox{\boldmath $i \:$}}
\newcommand{\Br}{\mbox{\boldmath $r \:$}}
\newcommand{\Bs}{\mbox{\boldmath $s \:$}}

\def\Beqa{\begin{eqnarray}}
\def\Eeqa{\end{eqnarray}}

\def\EX{{\rm e}}

\def\VEP{\varepsilon}

\title{%
Band Calculation for Ce-compounds on the basis of Dynamical 
Mean Field Theory
}

\author{%
Osamu \textsc{Sakai}\thanks{E-mail: sakai@phys.metro-u.ac.jp.},
Yukihiro \textsc{Shimizu}$^{1}$
and Yasunori \textsc{Kaneta}$^{2}$
}

\inst{%
Department of Physics, 
Tokyo Metropolitan University, Hachioji 192-0397, Japan \\
$^{1}$Department of Applied Physics, Tohoku University, Sendai 980-8579,
Japan \\
$^{2}$Department of Quantum Engineering and System Science, The
University of Tokyo,
Tokyo 153-8904, Japan
}

\recdate{
}
 
\abst{%
The band calculation scheme for $f$ electron compounds is 
developed on the basis of the
dynamical mean field theory (DMFT) and the LMTO method.
The auxiliary impurity problem is solved by a method named as 
NCA$f^{2}$v', which includes 
the correct exchange process of the $f^{1} \rightarrow f^{2}$
virtual excitation as the vertex correction 
to the non-crossing approximation (NCA)
for the $f^{1} \rightarrow f^{0}$ fluctuation.
This method leads to the correct magnitude of the Kondo temperature, $T_{\rm K}$, 
and makes it possible to carry out quantitative DMFT calculation including 
the crystalline field (CF) and the spin-orbit (SO) splitting of 
the self-energy. 
The magnetic excitation spectra are also calculated to estimate $T_{\rm K}$.
It is applied to Ce metal and CeSb at $T=300$ K as the first step.
In Ce metal, 
the hybridization intensity (HI) just below the Fermi energy 
is reduced in the DMFT band. 
The 
photo-emission spectra (PES) have a  conspicuous 
SO side peak, similar to that of experiments.
$T_{\rm K}$ is estimated 
to be about 70 K in $\gamma$-Ce, while to be about 1700 K
in $\alpha$-Ce.
In CeSb, the double-peak-like structure of PES
is reproduced.
In addition,  $T_{\rm K}$ which is not so low is obtained because
HI is enhanced  just at the Fermi energy in the DMFT band.
}

\kword{%}
dynamical mean field theory, band theory,
Ce metal, Ce-pnictides
}

\begin{document}
\sloppy
\maketitle

\section{Introduction}

The non-empirical band calculation on the basis of the local density
approximation (LDA) theory is one of the fundamental approaches
to obtain the insight into the electronic properties of solid state 
materials.
%It is successfully applied to predict physical properties in various
%research fields.
However, it has limitations to treat the dynamical excitation spectra
 (DES)
of the strongly correlated electron systems~\cite{B0}
such as $f$ electron compounds.
The understanding of DES of strongly
correlated electrons has been extensively improved recently based on the
approach of the dynamical mean field
theory (DMFT)~\cite{C1,A1}.
It seems natural to develop a non-empirical band calculation 
scheme for actual materials by the combination of 
LDA and DMFT~\cite{A2,A3,A4}.

In the DMFT, the strongly correlated band electron problem is
mapped onto the calculation of DES
of the 
auxiliary impurity Anderson model in an effective medium~\cite{A1}.
At present, however, we do  not have the theoretical method which
analytically gives correct DES of the impurity 
Anderson model.
Several numerical methods have been developed to calculate
DES of the Kondo problem.
It is known that the quantum Monte Carlo (QMC) 
method~\cite{A5,A6} and the numerical 
renormalization group (NRG) 
method~\cite{A7} give in 
principle correct DES.
They are applied to DMFT calculation~\cite{A6,A10},
but they have 
difficulties in 
application to the realistic band calculation 
of $f$ electron systems~\cite{A4,A14}.

Approximate but highly flexible approaches have been developed on the
basis of the resolvent 
technique~\cite{C7}.
The NCA method~\cite{C8,C10} has been  widely applied to various models 
though it has some weakness in application at very low temperatures:
it does not fulfill the Fermi liquid relation.
But it can be easily applied 
to realistic situations such as 
the competition between the Kondo effect and 
the crystalline field (CF) splitting~\cite{C10}.
NCA has been  also applied to the DMFT 
calculation~\cite{A20}.
The NCA equation is 
justified by the $1/N_{f}$ expansion
when the fluctuation of 
the valence is 
restricted to the $f^{1}$ and $f^{0}$ configurations, 
where $N_{f}$ is the degeneracy factor of the $f$ 
orbital~\cite{C10}.
However, the exchange coupling through the virtual excitation to 
the $f^{2}$
configuration is not  negligible in quantitative calculation.
A scheme which includes the $f^{2}$ configuration in the frame 
work of the NCA type diagram has been used.
However, it does not lead to the Kondo temperature ($T_{\rm K}$) given by 
using the  
exchange constant obtained by the Schrieffer-Wolff (S-W) 
transformation~\cite{C13}. 
Therefore, 
the exchange process through the $f^{2}$
configuration is not properly accounted in the scheme.
Actually, the DMFT calculation by using this
method gives a
too small Kondo resonance peak compared with the result using QMC
method~\cite{A4}.
%In the empirical approaches, the hybridization strength can be treated 
%as a fitting parameter to give the observed Kondo 
%temperature in analysis 
%of experiment in the frame work of NCA calculation.
%It works effectively because principal physical properties are scaled 
%by the Kondo temperature.
%However in the DMFT band calculation, it is desirable to use a 
%calculation method which gives correct order of magnitude of the 
%Kondo temperature since the hybridization strength itself is also 
%estimated by the band calculation.
In the DMFT band calculation, it is desirable to use a 
method which gives the correct order of magnitude of $T_{\rm K}$.

A method to include the $f^{2}$ configuration 
within the $1/N_{f}$ order has been 
developed~\cite{A16,C11,F1}.
The NCA equation is modified  by a vertex correction term which 
leads to exchange coupling through the $f^{2}$ configuration.
The self-consistency equation of this 
method needs huge computational time,
if one tries to solve it strictly.
But we can usually apply simplifying approximations for the 
equation~\cite{A16}.
It does not need so much computational time, only few times larger 
than that of the NCA calculation,
and gives $T_{\rm K}$ consistent with the result of 
the S-W transformation.
We call the method as NCA$f^{2}$v', henceforth, and is explained in 
Appendix.

%And the calculation of main diagonal part of the spectra 
%also does  not need so much computational time.
%But it needs very large computational time, about  order of magnitude 
%larger than  that of the main part,
%to calculate the  miner correction parts of the spectra due to the 
%off-diagonal contribution.
%The off-diagonal parts correspond to the interference effect of 
%the $f^{0} \leftrightarrow f^{1}$ and  
%$f^{1} \leftrightarrow f^{2}$  process in the single particle 
%final  states.
%In usual cases, we can safely neglect the time consuming correction 
%part.
%We call the method as NCA$f^{2}$v' method hereafter.
%It may be worthwhile to implement the NCA$f^{2}$v' method in DMFT band
%calculation.

Several efficient band calculation techniques have been 
developed~\cite{A17}.
Among them the LMTO method is a very convenient one because it has 
formal similarity to the usual LCAO picture~\cite{A18}.
It is widely used though accuracy is not so very high when it is 
compared with  other sophisticated methods.
In this paper we develop a DMFT band calculation scheme by the 
combination of NCA$f^{2}$v' and LMTO methods, and apply it to 
Ce metal and CeSb as the first step.

The effective hybridization intensity (HI)
is defined nominally as the 
product of the square of 
the hybridization matrix and the density of states of conduction 
bands in the auxiliary impurity problem.
It plays the key roles in the DMFT calculation.
We mainly study how HI is modified in the DMFT calculation from 
that of LDA theory, 
and 
how the effects of  
the spin-orbit (SO) splitting and the CF
splitting appear.

The relative occupation number of excited SO level is strongly
reduced  from that of LDA  in the correlated bands 
even when the total occupation number of the $4f$ electron is fixed near 
the value of LDA result.

Ce metal is known to have iso-structural {\it fcc} 
phases~\cite{A19}.
The low temperature state is called as the $\alpha$ phase and it has 
smaller lattice constant.
The higher temperature one is called as the $\gamma$ phase.
The DMFT studies on this transition have been extensively 
carried out~\cite{A4,A20,A21,A22}.
In these studies the SO and the CF splittings were neglected.
In the present study we are not concerned with the $\alpha-\gamma$ 
transition, but we calculate the photo-emission spectra 
(PES)~\cite{B1,A27,A23,A24}
for the $fcc$ state with the lattice constant of these two 
cases.
The HI estimated by the LDA band has a peak just below $E_{\rm F}$.
This peak is reduced in the DMFT band, thus the SO side peak in PES becomes
relatively conspicuous.

The CF level splitting is small in the auxiliary impurity model 
of $\gamma$-Ce, 
and $T_{\rm K}$ is estimated to be about 70 K by the calculation of the
magnetic excitation.
In $\alpha$-Ce, $T_{\rm K}$ of the auxiliary impurity model 
becomes very high about 1700 K.

%The energy level of $(j=5/2)\Gamma_{8}$ of the auxiliary impurity model 
%is lower about 20 K than the energy 
%of the $(j=5/2)\Gamma_{7}$ in $\gamma$ Ce, while
%$T_{\rm K}$ in the lowest $(j=5/2)\Gamma_{8}$ states is estimated 
%to be about 50 K from the magnetic excitation.
%They have  comparable magnitude in $\gamma$ state. 
%The CF splitting and $T_{\rm K}$ become, respectively,  180 K and 320 K
%in  $\alpha$-Ce, both are large, but  also have comparable magnitude.

CeSb would be a first material for which HI 
calculated by the realistic band is used in the 
analysis of PES~\cite{A27,A26,A29}.
The hybridization matrix  is not small in the valence band region
of Sb $p$ states.
In the energy region near $E_{\rm F}$ 
the density of states of bands is small
reflecting the semi-metallic nature of the band structures.
Various anomalous properties such as the very large magnetic 
anisotropy and  the origin of the 
ferromagnetic layers in the ordered state are explained 
on the basis of the $p-f$ mixing model proposed
by Kasuya {\it et. al.}~\cite{A30,A31,A32,A33}
The top of the valence  $p$ band has the $(j=3/2)\Gamma_{8}$ symmetry.
A part of  $\Gamma_{8}$ bands 
which has the same symmetry with that of the occupied 4f band
is pushed up above 
$E_{\rm F}$  due to the $p-f$ hybridization
in the 
ordered phase,
and this plays important roles in the $p-f$ mixing model.
Such re-constructed electronic band structures in the ordered states 
are confirmed by the dHvA 
and the optical 
experiments~\cite{A34,A36,A40}.
In the ordered states we can use the Hatree-Fock-like static mean 
field approaches, and detailed comparison with experiments has been 
carried 
out~\cite{A33,A45}.
However, in the paramagnetic phase we need DMFT band calculation.

CeSb has a characteristic double peak structure in PES; 
the shallow-energy one is at about 
0.8 eV below $E_{\rm F}$ and the deep-energy one is at about 3 eV below 
$E_{\rm F}$~\cite{B1,A29,A49,D1}.
This has been explained by the large  peak of HI in the valence 
band region~\cite{A27,A26,A29}.
$T_{\rm K}$ has been calculated by using the same HI
on the basis of the impurity model\cite{A51,A54}.
It is very low, about $10^{-5}$ K 
because of the small density of states
near $E_{\rm F}$.
However  higher $T_{\rm K}$, about 10 K, has been expected  from
the transport properties~\cite{A31}.
In the present DMFT calculation, the double-peak-like structure is 
obtained.
At the same time we obtained a rather higher $T_{\rm K}$, 
since HI of DMFT has a sharp peak just at $E_{\rm F}$
caused by  the hybridization of the  bands 
with the correlated $4f$ bands.
Similar evolution of the peak at $E_{\rm F}$ in HI has been observed
recently in 
DMFT calculation for CeP and CeAs in ref. \citen{D2}.
But the 
SO splitting and the correct exchange process through the  $f^{2}$ 
state have not been  included there.
The $(j=5/2)\Gamma_{8}$ state has  lower energy than that of 
$(j=5/2)\Gamma_{7}$ in the present calculation starting directly from
the LDA band.
It is demonstrated that more
similar result with experiments such as the definite double peak structure 
of PES and
lower, but not so low, $T_{\rm K}$
can be obtained when 
the band overlapping between 
the occupied Sb $p$ bands and the un-occupied Ce $d$ bands is reduced 
from that of the LDA band.

In both materials,
we find that 
HI which is directly calculated from the $4f$ partial DOS of LDA bands 
seems to be somewhat large 
when we compare the calculated PES
with those of experimental ones~\cite{B1,A23,A51}.
But in the DMFT calculation the effect of HI is relatively 
weakened.
The calculated HI is somewhat lager, but not so drastically large 
than the magnitude expected from experimental data.

In \S 2, we give our formulation on the basis of the LMTO method.
The DES which are calculated by the NCA$f^{2}$v' method for 
the single impurity model are 
compared with the results of NRG in \S 2.
Applications to Ce metal with the lattice constant of $\gamma$-Ce, 
 and  with that of $\alpha$-Ce are given in \S 3 and \S 4.
In \S 4 the calculation for CeSb is shown.
Summary is given in \S 6.
In Appendix we explained the NCA$f^{2}$v' method.

\section{Formulation
}

\subsection{LMTO-DMFT Matrix Equation}

We consider the spectra  of the following Hamiltonian~\cite{A4},
\Beqa
{\cal H}={\cal H}_{\rm LDA} 
+
\frac{U}{2}\sum_{\Bi}
(\sum_{\Gamma,\gamma}
c^{+}_{\phi^{\rm a}\Bi\Gamma\gamma}
     c_{\phi^{\rm a}\Bi\Gamma\gamma}
 -n^{\rm LDA}_{\Bi f})^{2}.
\label{eq.2.1}
\Eeqa
Here 
$ c_{\phi^{\rm a}\Bi\Gamma\gamma}$ is the annihilation 
operator for the atomic localized state,
$
\phi^{\rm a}_{\Bi\Gamma\gamma}(\Br)
$,
at site  $\Bi$ with  the 
$\gamma$  orbital of the $\Gamma$ irreducible representation. 
The quantity $n^{\rm LDA}_{\Bi f} = 
\sum_{\Gamma\gamma}n^{\rm LDA}_{\phi^{\rm a}\Bi\Gamma\gamma}$ 
is the 
occupation number on the atomic $4f$-electron per a Ce ion in 
the LDA calculation.
We assume that the local Coulomb interaction is expressed by the 
$\phi^{\rm a}_{\Bi\Gamma\gamma}$ orbit.

The excitation spectra are expressed by introducing the self energy
terms~\cite{A1},
\Beqa
{\cal H}_{\rm DMFT}={\cal H}_{\rm LDA} 
+ \sum_{\Bi,(\Gamma,\gamma)}
(\Sigma_{\Gamma}(\VEP+{\rm i}\delta)
 +\VEP^{\rm a}_{\Gamma}-\VEP_{\Gamma}^{\rm LDA})
 |\phi^{\rm a}_{\Bi\Gamma\gamma} >
     < \phi^{\rm a}_{\Bi\Gamma\gamma} |,
\label{eq.2.2}
\Eeqa
where $\VEP^{\rm a}_{\Gamma}$ is the single electron energy level of 
$4f$ state, and 
$\VEP_{\Gamma}^{\rm LDA}$ is the energy level 
in the  LDA calculation.
The self-energy $\Sigma_{\Gamma}(\VEP+{\rm i}\delta)$ is calculated by solving the 
auxiliary impurity problem in the effective medium with the use of  the 
NCA$f^{2}$v' method~\cite{A16}, the outline of it is described in Appendix.

Usually, the DMFT calculation is carried
out after transforming 
the ${\cal H}_{\rm LDA}$ part into the LCAO scheme or to the localized 
Wannier representation, and then using those basis functions~\cite{A4}.
In this paper we include the self-energy term directly in the LMTO 
matrix.
This simplifies the calculation because the processes of defining the 
localized basis function are skipped, but instead, 
Greenian equation in the non-orthogonalized bases must be 
introduced.
In later calculation we will approximate the localized $4f$ state
$\phi^{\rm a}$
 by the band center orbit
$\phi(-)$ which has 
logarithmic derivative $-\ell-1$ on the muffin-tin surface, 
because it it well localized for the $4f$ state.

In the LMTO method~\cite{A18}, the Hamiltonian 
${\cal H}_{\rm LDA}$ is represented as the matrix 
using the LMTO bases,
\Beqa
\psi^{j\Bk}(\Br)
=\sum_{q L}a^{j\Bk}_{q L}
\chi^{\Bk}_{q L}(\Br).
\label{eq.2.3}
\Eeqa
Here, $a^{j\Bk}_{q L}$ is the 
expansion coefficient 
of the $j$-th eigen state with the wave number $\Bk$
on the LMTO base of the angular momentum
($\ell,m$) and spin ($\alpha$) 
 at site $q$ 
in the unit cell.
We denote as $L \equiv (\ell,m,\alpha)$, henceforth.
The LMTO Bloch state is given as 
\Beqa
\chi^{\Bk}_{q L}(\Br)
=\frac{
\Phi_{q L}(-;\Br_{q})
}
{
\sqrt{S_{q}/2}\Phi_{q\ell}(-)}
\delta_{qq'}
-\sum_{L'}
\frac{
\Phi_{q' L'}(+;\Br_{q'})
}
{
(2\ell'+1)\sqrt{S_{q'}/2}\Phi_{q'\ell'}(+)
}
S^{\Bk}_{q' L', q L},
\label{eq.2.4}
\Eeqa
in the sphere of $q'$, 
where $\Br_{q'} = \Br-\BR_{q'}$ is the electron coordinate from the center
of MT at $\BR_{q'}$.
The quantity $S_{q}$ is the radius of the MT sphere, and
$S^{\Bk}_{q' L', q L}$ is the structure factor~\cite{A18}.
The LMTO base function with logarithmic derivative $D$ is given 
\Beqa
\Phi_{q L}(D;\Br)
={\rm i}^{\ell}Y^{m}_{\ell}(\hat{\Br}_{q})\chi(\alpha) 
(\phi_{\nu\ell}(r_{q})
+\omega(D)\dot{\phi}_{\nu\ell}(r_{q})),
\label{eq.2.5}
\Eeqa
with the omega factor $\omega(D)$  of Andersen~\cite{A17}.
The quantity
$\Phi_{q L}(-;\Br_{q})$ ($\Phi_{q L}(+;\Br_{q})$)
 in eq. (\ref{eq.2.4}) means 
$\Phi_{q L}(-(\ell+1);\Br_{q})$ ($\Phi_{q L}(\ell;\Br_{q})$),
and 
$\Phi_{q \ell}(\pm)$ is
$
\phi_{\nu\ell}(r_{q})+\omega(\pm)\dot{\phi}_{\nu\ell}(r_{q})
$ on the MT sphere.
The wave function $\phi_{\nu\ell}(r_{q})$ and its  
dot state, $\dot{\phi}_{\nu\ell}(r_{q})$
are calculated at the energy $\VEP_{\nu}$. 

The density of states (DOS) of the excitation is obtained by 
solving the Greenian equation of the matrix form,
\Beqa
[ z\hat{O}-\hat{{\cal H}}_{\rm LDA}-\hat{\Sigma}(z)]\hat{G}(z) =
\hat{I}.
\label{eq.2.6}
\Eeqa
The notation hat of the matrices means that they are
defined on the bases of $\chi^{\Bk}_{q L}$.
As noted previously, the LMTO bases are  not orthogonalized 
with each others~\cite{A18}.
Therefore, we have the overlapping integral $\hat{O}$ in
eq. (\ref{eq.2.6}),
whose matrix element is given as $O^{\Bk}_{q'L',qL}
=(\chi^{\Bk}_{q'L'} \mid \chi^{\Bk}_{qL}) $.
The DOS on the $f$-state is given by 
\Beqa
 \rho^{({\rm band})}_{\Gamma}(\VEP)=-\frac{1}{\pi}{\rm tr}
[\hat{O}_{\Gamma}\hat{G}(\VEP+{\rm i}\delta)].
\label{eq.2.7}
\Eeqa
The projection operator is defined as
\Beqa 
\hat{O}_{\Gamma}=\sum_{\Bi\gamma}
|\phi^{\rm a}_{\Bi(\Gamma\gamma)}><\phi^{\rm a}_{\Bi(\Gamma\gamma)}|.
\label{eq.2.8}
\Eeqa

The DMFT self-consistency calculation is carried out by the following 
way.
(I) At first we solve the auxiliary impurity model with a trial HI and 
impurity level $\VEP^{({\rm imp.})}_{\Gamma}$, 
and calculate the DOS $\rho^{\rm imp.}_{\Gamma}(\VEP)$, 
then calculate the local impurity Green's function by the Cauchy 
integral,
\Beqa
G^{({\rm imp.})}_{\Gamma}(z)=\frac{1}{g_{\Gamma}}
\int {\rm d}x \frac{\rho^{({\rm imp.})}_{\Gamma}(x)}{z-x},
\label{eq.2.9}
\Eeqa
where $g_{\Gamma}$ is the degeneracy factor of the $\Gamma$ symmetry 
state.
(II) Then the
electron self-energy due to the Coulomb interaction,
$\Sigma_{\Gamma}(z)$ is calculated by the relation,
\Beqa
\Sigma_{\Gamma}(z)
=z-\VEP^{({\rm imp.})}_{\Gamma}-G^{({\rm imp.})}_{\Gamma}(z)^{-1}
-\Sigma^{({\rm h})}_{\Gamma}(z).
\label{eq.2.10}
\Eeqa
Here $\Sigma^{({\rm h})}_{\Gamma}(z)$ is the hybridization self-energy of 
the impurity problem, which is calculated by the Cauchy 
integral from the effective trial HI~\cite{A55}.
(III) By using the self-energy $\Sigma_{\Gamma}(z)$ in (\ref{eq.2.6}),
the  band problem is solved, and using the equation (\ref{eq.2.7}), 
the DOS of the  band 
$\rho^{({\rm band})}_{\Gamma}(\VEP)$
is calculated.
(IV) If $\rho^{({\rm band})}_{\rm \Gamma}(\VEP)$ and 
$\rho^{({\rm imp.})}_{\rm \Gamma}(\VEP)$
agree with each other within the tolerance, the calculation is 
stopped. 
(V) If both do not agree, local band Green's function, 
$G^{({\rm band})}_{\Gamma}(z)$ is calculated by 
the Cauchy integral, and we estimate new effective hybridization
self-energy from the relation,
$
\Sigma^{({\rm h:new})}_{\Gamma}(z)
=z-\VEP^{({\rm eff.})}_{\Gamma}-G^{({\rm band})}_{\Gamma}(z)^{-1}
-\Sigma_{\Gamma}(z)
$,
where $\VEP^{({\rm eff.})}_{\Gamma}$ is the first moment 
of $\rho^{({\rm band})}_{\Gamma}(\VEP)$.
(VI) Then  new effective HI is calculated from the 
imaginary part
of $\Sigma^{({\rm h:new})}_{\Gamma}(\VEP+{\rm i}\delta)$.
We return to the step (I).

We note that the spectrum intensity in the resolvent method is 
not usually normalized to be unity, because
configurations of the $4f$ state are restricted to some important 
ones~\cite{C10}.
This  is overcome by introducing the normalization factor 
$Z_{\Gamma}$ and the level shift $\Delta\VEP_{\Gamma}$ as
\Beqa
\Sigma_{\Gamma}(z)
=\Delta\VEP_{\Gamma}+
\tilde{\Sigma}_{\Gamma}(z)-(Z^{-1}_{\Gamma}-1)(z-E_{\rm F}),
\label{eq.2.11}
\Eeqa
where  $1/Z^{-1}_{\Gamma}$ is  
the total integrated intensity of DES,
and $\Delta\VEP_{\Gamma}$ is determined from the first 
moment of $\rho^{({\rm imp.})}_{\rm \Gamma}(\VEP)$.
These ensure
the relation 
$\tilde{\Sigma}_{\rm \Gamma}(z) \sim O(1/z)$ for large $|z|$,
which is required from the usual Kramers-Kronig relation.
This means that the eq. (\ref{eq.2.9})  is expressed as 
$G^{({\rm imp})}_{\Gamma}(z)
=[
z-\VEP^{({\rm imp.})}_{\Gamma}-\Delta\VEP_{\Gamma}
-\tilde{\Sigma}_{\Gamma}(z)
+(Z^{-1}_{\Gamma}-1)(z-E_{\rm F})
-\Sigma^{({\rm h})}_{\Gamma}(z) 
]^{-1}
$.

\subsection{NCA$f^{2}$v' Method
}

\begin{figure}[htb]
\begin{center}
\includegraphics[width=8cm,clip]{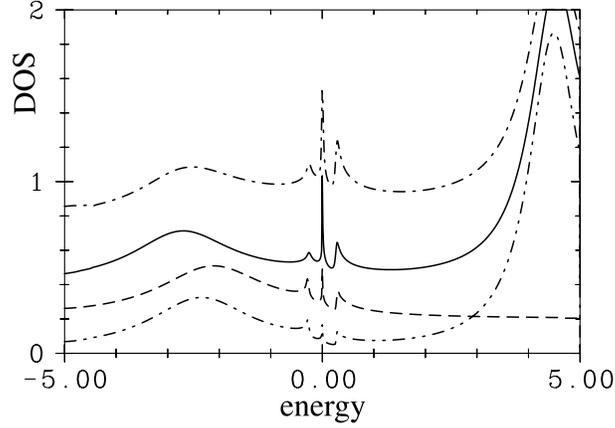}
\end{center}
\caption{Comparison of spectra calculated by several methods on the
 basis of 
the resolvent technique.
The dashed line is the result of usual NCA.
The two-dots-dashed line is given by NCA type diagram including 
the $f^{2}$ states. The solid line  is given by NCA$f^{2}$v' and the 
dot-dashed line is given by NCA$f^{2}$v.
Two doublets are located at $\VEP_{f1}=-2.2$ and $\VEP_{f2}=-1.9$.
The Coulomb constant is $U=6$ and the hybridization
intensity is $v^{2}\rho_{c}=0.08$, the  
density of states of band is  constant between ($-D \sim D$)
with $ D=5$. $T=1.723 \times 10^{-3}$. The DOS of this figure is
 normalized as the states/spin.
The zero of the lines are shifted by 0.2 successively, but the dot-dashed 
 line by 0.4 from that of the solid line.
}
\label{fig:1}
\end{figure}

Before going to the DMFT calculation, we compare briefly the 
solutions to the impurity problem obtained by four typical methods 
of resolvent 
technique~\cite{A16}.
By the solid line in Fig.~\ref{fig:1}, we show the single particle 
spectra
calculated by the NCA$f^{2}$v' method for simplified 
hybridization. 
We have assumed that: the conduction band has constant density of 
states between the energy region $(-D \sim D)$ with $D=5$;
HI is $v^{2}\rho_{\rm c}=0.08$
with the hybridization matrix $v$ and the density of states of
the conduction band $\rho_{\rm c}$;
the doubly degenerate $f$-states are located at 
$\epsilon_{f1}=-2.2$ and $\epsilon_{f2}=-1.9$;
the Coulomb interaction constant between the $f$-electrons is $U=6$.
The temperature is set at 
 $T=1.723 \times 10^{-3}$.
There appears the Kondo resonance peak just at the Fermi energy, 
and satellites on both sides of it.
The satellites correspond to the 
shake up excitation $\epsilon_{f1} \rightarrow \epsilon_{f2}$ in 
the creation of hole or electron near the Fermi energy.
The result obtained by usual NCA for the ($f^{0}$-$f^{1}$) fluctuation model 
is given by the dashed line.
The Kondo resonance peak is small because the exchange coupling 
through the $f^{2}$ configuration is not considered.
Grossly $T_{\rm K}$ is estimated as
$D\EX^{-\frac{1}{N_{f}J\rho_{\rm c}}}$ 
with 
the exchange interaction constant, $J$.
The constant $N_{f}J\rho_{\rm c}$ is given  as 
$N_{f}v^{2}\rho_{\rm c}/|\epsilon_{f}|$
in the usual NCA,
where $|\epsilon_{f}|$ is the excitation energy of 
$f^{1} \rightarrow f^{0}$.
When the virtual excitation  to the $f^{2}$ states is included 
in the S-W transformation~\cite{C13},  
this factor is replaced 
by the expression
$N_{f}v^{2}\rho_{\rm c}/|\epsilon_{f}|
+N_{f}v^{2}\rho_{\rm c}/|\epsilon_{f}+U|
$.
The quantity
$| \epsilon_{f} +U |$ is the excitation energy of
$f^{1} \rightarrow f^{2}$.
Though  we usually expect the relation  
$|\epsilon_{f}| < |\epsilon_{f}+U|$,
the second factor gives a comparable magnitude with the first term.
Because  the inverse  of $J$ appears in the exponential function,
the small increase of $J$ greatly increases $T_{\rm K}$.

%Fig1

If we tentatively use $N_{f}=2$ by assuming the energy difference
$\epsilon_{f2}-\epsilon_{f1}$ to be large, 
$T_{\rm K}$
is estimated as 
$T_{\rm K, S-W} =7.0 \times 10^{-4}$
and 
$T_{\rm K, NCA} =5.3 \times 10^{-6}$.
If we use $N_{f}=4$ by assuming that the energy 
difference is small, $T_{\rm K}$ is 
estimated as 
$T_{\rm K, S-W} =0.073$
and 
$T_{\rm K, NCA} = 0.0085$.
These estimated values are largely different with each other.
In quantitative calculations, 
it is important to calculate the DES
by including the $f^{2}$ configuration, and also the SO
and CF energy splitting.

The spectra obtained by the sum of the NCA type diagrams for the 
($f^{0}$-$f^{1}$-$f^{2}$) fluctuation model is given by the
two-dots-dashed line.
The Kondo resonance peak in this method is smaller than
that of the usual 
 ($f^{0}$-$f^{1}$) type NCA calculation.
We think this result is caused by the relative decrease in $f^{1}$ state 
energy:
the virtual excitation to the $f^{2}$ state 
has order $N_{f}-1$ effect, and thus decreases energy in $f^{1}$ 
state largely;
this
decreases the exchange coupling of the virtual 
excitation to the $f^{0}$ state.
On the other hand the exchange coupling through the virtual 
excitation to the $f^{2}$ state  is not included 
in this approximation.
Therefore $T_{\rm K}$ decreases.

We show by the dot-dashed line the spectra
calculated by a method named as NCA$f^{2}$v.
Details of which are explained in the Appendix.
In this method, the off-diagonal contribution of the vertex 
is added to the spectra of NCA$f^{2}$v'.
It gives the interference effect 
of $f^{0} \leftrightarrow f^{1}$ and 
$f^{1} \leftrightarrow f^{2}$ processes in the final state of the 
single particle excited states.
This effect transfers some intensity from the high excitation energy 
region to the low energy region,
but total  intensity is not changed~\cite{A16}.
The correction is not so drastic when compared with
the change from NCA to NCA$f^{2}$v'.
The calculation of this off-diagonal correction needs the very large 
computing time of order
$N_{\VEP}^{3}$, while the time for diagonal part 
needs only $N_{\VEP}^{2}$, where 
$N_{\VEP}$ is the number of mesh points of energy.
$N_{\VEP} \sim 10^{5}$ is needed because the width of 
the energy mesh should be smaller than the Kondo temperature.
We use the NCA$f^{2}$v' method in the 
DMFT band calculation in this paper.

\begin{figure}[htb]
\begin{center}
\includegraphics[width=8cm,clip]{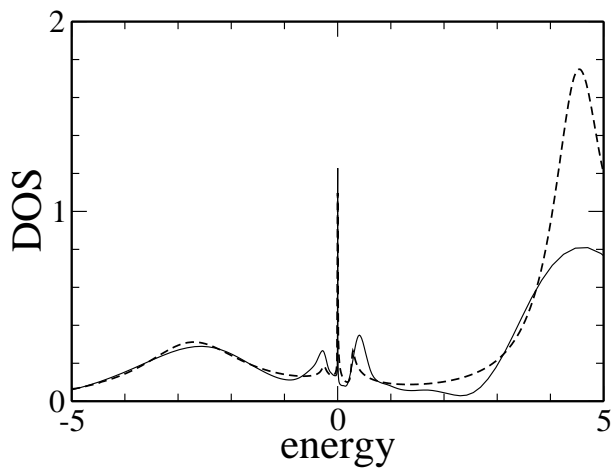}
\end{center}
\caption{Comparison of spectra calculated by  NCA$f^{2}$v'(dashed
 line)
 and  NRG (solid line) methods. 
The temperatures are $T=0.7 \times 10^{-4}$ in NCA$f^{2}$v' and $T=0$ in NRG. 
The parameters are the same as those in 
Fig.~\ref{fig:1}. The discretization parameter of NRG is $\Lambda=2$.
The peak height at the Fermi energy of NCA$f^{2}$v' method is 
about $1.1$.
}
\label{fig:2}
\end{figure}

%Fig2

In Fig.~\ref{fig:2}, we compare the spectra obtained by NRG 
method~\cite{A7} (solid line) at $T=0$ and the 
NCA$f^{2}$v' method (dashed line) at $T=0.7 \times 10^{-4}$.
This temperature is  about 1/10 of the Kondo temperature, 
$T_{\rm K,SW}(N_{f}=2)=7 \times 10^{-4}$, 
and will be  low enough to compare the result of NRG at $T=0$.
The intensities of the peaks at  the Fermi energy are comparable, 
slightly small in the NCA$f^{2}$v' method.
The integrated intensity of the shake up  side peaks is 
considerably small in the 
NCA$f^{2}$v'.
This points will be improved if we use the NCA$f^{2}$v method, 
but we restrict ourselves to the NCA$f^{2}$v' in this paper.
We note, even the intensity of the shake up peaks is small, the present
method gives the energy scale of Kondo effect not so different from the 
calculation of NRG.
Usual NCA and NCA$f^{2}$ give some order of magnitude lower 
Kondo temperature.

\section{ $\gamma$-Ce
 }

\subsection{
The hybridization intensity of  LDA band 
}
\begin{figure}[htb]
\begin{center}
\includegraphics[width=8cm,clip]{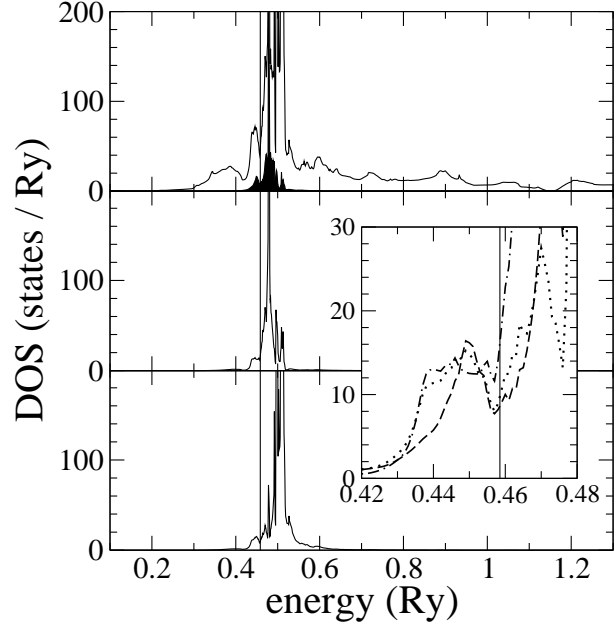}
\end{center}
\caption{The density of states (DOS) obtained by LDA band calculation
for $\gamma$-Ce.
The solid line in the top panel
is the total DOS, 
the shaded region 
is DOS on the $(j=5/2)\Gamma_{7}$, 
the middle panel 
is  DOS on the $(j=5/2)\Gamma_{8}$, and
the bottom panel
is  DOS on the $j=7/2$ states.
The Fermi energy $E_{\rm F}=0.4585$ Ry is indicated by the vertical 
solid  line.
In the inset, DOS's near $E_{\rm F}$ are shown 
for 
$(j=5/2)\Gamma_{7}$(dashed line), $(j=5/2)\Gamma_{8}$(dot-dashed line)
and 
$j=7/2$(dotted line) states.
The standard LMTO method with ASA and CC described  in ref. \citen{A18} 
is used.
The exchange correlation function  of Gunnarsson-Lundqvist~\cite{A56} 
is employed
 and the lattice constant is $a=9.75477$ AU.
The frozen core approximation is used  for $5p$ states.
}
\label{fig:3}
\end{figure}

As noted previously we approximate $\phi^{\rm a}_{\Bi\Gamma\gamma}$
by  $\phi_{\Bi\Gamma\gamma}(-)$,
and $\VEP_{\nu}=\VEP^{({\rm LDA})}_{\Gamma}$ is chosen to fulfill the condition
$\omega(-)=0$.
The energy $\VEP^{\rm a}_{\Gamma}=\VEP^{\rm a}_{4f}$ is 
changed as a parameter
to lead the target of the total $4f$ occupation number.
In Fig.~\ref{fig:3},  we show DOS
calculated by the LDA 
band for 
the lattice constant of 
$\gamma$-Ce
($a= 9.75477$ AU).
The $5p$ of Ce is treated as the frozen core state
and the scalar relativistic approximation is used.
The mesh of k-points in the band calculation is generated by taking 
9 points on the  $\Gamma - X$ axis (including $\Gamma$ and X points). 
The SO interaction 
is added as 
$\zeta_{\ell}\Bs\cdot\Bell$ in the matrix.
The solid line in the top panel
gives the total DOS.
The shaded region in the panel
gives the partial DOS on the $(j=5/2)\Gamma_{7}$, 
the middle panel 
gives the partial DOS on the $(j=5/2)\Gamma_{8}$, and
the bottom panel
gives partial DOS on the $j=7/2$ state.
The Fermi energy $E_{\rm F}=0.4585$ Ry is indicated by the vertical 
solid  line.
For the $j=7/2$ group, DOS is summed over CF states
because 
their splitting is not so important since 
they have higher excitation energy in the DMFT band.
In the inset, the partial DOS's near $E_{\rm F}$ are shown 
for the 
$(j=5/2)\Gamma_{7}$(dashed line), $(j=5/2)\Gamma_{8}$(dot-dashed line) 
and 
$j=7/2$(dotted line) states.

%Fig3

The total occupation number of the $4f$  electron is obtained as $1.0859$,
and partial occupation numbers are shown in Table \ref{t1}.
Hereafter we denote 
$\phi_{\Bi\Gamma\gamma}(-)$ as  
$\phi_{\Bi\Gamma\gamma}$.
By using DOS 
 $\rho^{({\rm LDA})}_{\Gamma}(\VEP)$
on the orbit
$\phi_{\Bi\Gamma\gamma}$,
we define  Green's function 
$G^{({\rm LDA})}_{\rm \Gamma}(z)$
by the Cauchy integral same to 
eq. (\ref{eq.2.9}).
%\Beqa
%G^{({\rm LDA})}_{\Gamma}(z)=
%\int {\rm d}x \frac{\rho^{({\rm LDA})}_{\Gamma}(x)}{z-x}.
%\Eeqa
Following ref. \citen{A55}, the hybridization self-energy
for the non-interacting case is 
given as 
$
\Sigma^{({\rm h,LDA})}_{\Gamma}(z)
=z-\VEP^{({\rm eff,LDA})}_{\Gamma}-G^{({\rm LDA})}_{\Gamma}(z)^{-1}.
$
%\Beqa
%\Sigma^{({\rm h,LDA})}_{\Gamma}(z)
%=z-\VEP_{\Gamma}-G^{({\rm LDA})}_{\Gamma}(z)^{-1}.
%\Eeqa
The HI is obtained from the imaginary part of 
$\Sigma^{({\rm h,LDA})}_{\Gamma}
(\VEP+{\rm i}\delta)$.

%Fig4
\begin{figure}[htb]
\begin{center}
\includegraphics[width=8cm,clip]{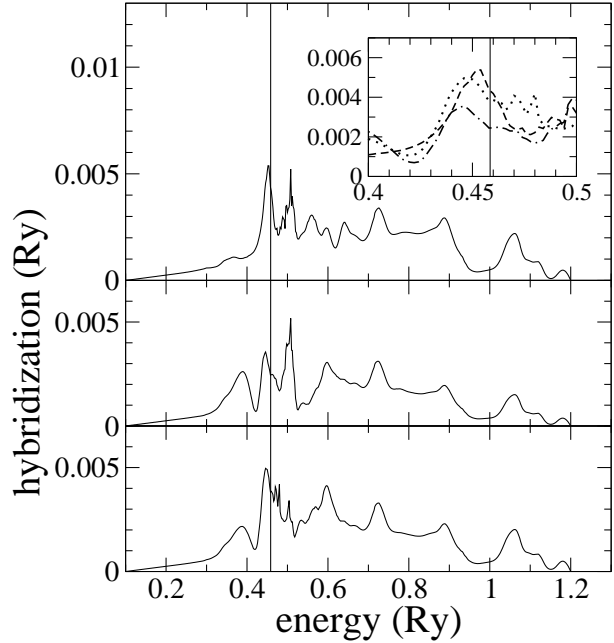}
\end{center}
\caption{Hybridization intensity (HI) calculated by the LDA band for 
$\gamma$-Ce.
The upper panel
is HI for $(j=5/2)\Gamma_{7}$, 
the middle panel 
is  one for $(j=5/2)\Gamma_{8}$, and
the lower panel
is  one for $j=7/2$ state.
In the inset, HI's near $E_{\rm F}$ are shown 
for 
$(j=5/2)\Gamma_{7}$(dashed line), $(j=5/2)\Gamma_{8}$(dot-dashed line) 
and 
$j=7/2$(dotted line) states.
Small HI is added below the band bottom (at $0.2029$ Ry), see the text.
}
\label{fig:4}
\end{figure}

The HI for the LDA band is shown in Fig.~\ref{fig:4}.
We note that HI of $(j=5/2)\Gamma_{7}$
(upper panel) is relatively
large near $E_{\rm F}$  where the $5d$ band has the large 
density of states, while 
it is small
in the energy region of the deep valence band where the  6s band is dominant.
In the inset of Fig.~\ref{fig:3}, we have shown DOS for 
$f$-components  near $E_{\rm F}$.
The DOS of $(j=5/2)\Gamma_{7}$  is 
relatively larger just below the Fermi energy
than that of the $(j=5/2)\Gamma_{8}$.
The degeneracy factor of the state is half of the $(j=5/2)\Gamma_{8}$.
Therefore, HI of the $(j=5/2)\Gamma_{7}$ is 
about twice of the $(j=5/2)\Gamma_{8}$ component even when DOS 
has comparable value.
This fact can be seen in Fig.~\ref{fig:4}.
The HI of the $j=7/2$ has the 
comparable magnitude with that of the $(j=5/2)\Gamma_{7}$ near 
$E_{\rm F}$, 
and with that of $(j=5/2)\Gamma_{8}$ in the deep region. 
The averaged HI over these three groups is similar to that used in 
ref. \citen{A24}.

When HI is strong, the $f^{0}$-like  peak 
in  DOS of the correlated band  
appears below the band bottom ($0.2020$ Ry).
In such case the peak becomes very sharp, thus 
it is not easy to proceed the  iteration steps.
To avoid such situation, we added small HI in the 
energy region below the band bottom in the self-consistent 
DMFT calculation~\cite{A57}.
The added part is included in Fig. 4.

%Fig5

\begin{figure}[htb]
\begin{center}
\includegraphics[width=8cm,clip]{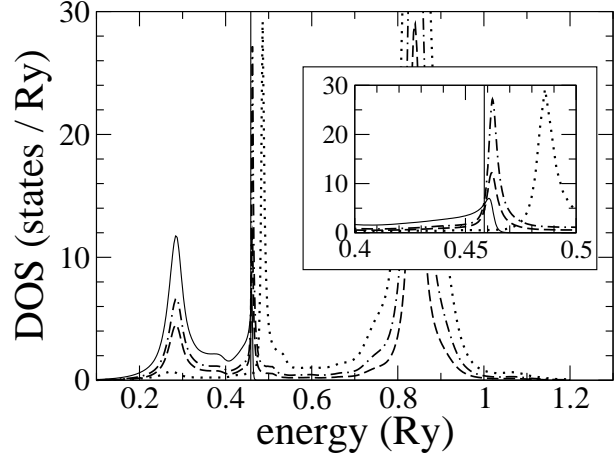}
\end{center}
\caption{$4f$ spectra of the impurity model by using HI obtained by the LDA band
for $\gamma$-Ce at $T=300$ K. 
The solid line is the total $4f$ PES spectra.
The dashed line is  DOS of $(j=5/2)\Gamma_{7}$,
the dot-dashed line is DOS of $(j=5/2)\Gamma_{8}$ and
the dotted-line is DOS of $j=7/2$ components.
Inset shows the spectra near the Fermi energy.
}
\label{fig:5}
\end{figure}

The $4f$ spectra calculated by using the NCA$f^{2}$v' method 
at $T=300$ K for this 
hybridization is shown in Fig.~\ref{fig:5}.
The levels
are chosen to be equal to those 
obtained in the self-consistent calculation.
The solid line gives the total $4f$ PES.
The single particle DOS's for each component is also shown.
Because the multiplet splitting of $4f^{2}$ states is neglected,
we have only single large peak at about $\VEP_{f}+U \sim 0.8$ Ry
for the $f^{1} \rightarrow f^{2}$ excitation.
The Coulomb interaction constant is assumed to be $U=6.5$ eV $( =0.48$ Ry).

%Table1
\begin{table}[t]
\caption{Various quantities obtained in DMFT calculation for $\gamma$-Ce.
$n_{\Gamma}^{\rm LDA}$ is the occupation number
in the LDA band,
$n_{\Gamma}^{({\rm imp'.})}$ is the occupation number
obtained by the impurity problem using HI of the LDA band.
$n_{\Gamma}^{({\rm band})}$ is the occupation number in the DMFT 
band.
$n_{\Gamma}^{({\rm imp.})}$ is the occupation number
in the auxiliary 
impurity problem of effective HI of DMFT calculation, 
 $\VEP_{\Gamma}$ is the energy level,
$\bar{Z}_{\Gamma}$ is the renormalization factor of the $f$ band,
$\bar{\VEP}_{\Gamma}$ is the effective energy of the renormalized band,
and $\bar{\Gamma}_{\Gamma}$ is the imaginary part of the self-energy.
The energy levels are measured from the Fermi energy 
$E_{\rm F}=0.4585$ Ry, and given in Ry. 
$\VEP^{\rm a}_{4f} = -0.12170$ Ry.
$\zeta_{4f}=0.007032$ Ry.
}
\label{t1}
\begin{halftabular}{@{\hspace{\tabcolsep}\extracolsep{\fill}}cccc} \hline
%\multicolumn{4}{c} {$r=1$} &\multicolumn{3}{c} {$ r=0.7$} \\
 & $\Gamma_{7}$ & $\Gamma_{8}$ & $j=7/2$
\\ \hline
$n_{\Gamma}^{\rm LDA}$      &
0.4017   & 0.2845   & 0.3997 \\
$n_{\Gamma}^{({\rm imp'.})}$  &
0.3396   & 0.5055   & 0.0949
                              \\
$n_{\Gamma}^{({\rm band})}$ & 
0.3296   & 0.6168   & 0.0364
                              \\
 $n_{\Gamma}^{({\rm imp.})}$ & 
 0.3292  & 0.6165   & 0.0371 
                              \\
$\VEP_{\Gamma}$                &
-0.11943  &-0.12016 &-0.09595
                               \\  
$\bar{Z}^{-1}_{\Gamma}$        & 
37.9     & 36.3  & 2.7
                               \\
$\bar{\VEP}_{\Gamma}$            &
0.15098  &0.15952 & 0.31193
                              \\ 
$\bar{\Gamma}_{\Gamma}$            &
0.1533  &0.1454 & 0.0002
                               \\
\hline
\end{halftabular}
\end{table}

The occupation numbers
are shown in the row of $n_{\rm \Gamma}^{({\rm imp'.})}$ in the 
Table \ref{t1}.
The occupation number of $j=7/2$ components
is
reduced to $0.0949$ from the value, $0.3997$ of the LDA calculation.
Even when the total occupation number of the $4f$ electron is not so different 
from the value of LDA, the partial occupation will be largely changed
in the strongly correlated bands.

The relative occupation of the $(j=5/2)\Gamma_{7}$ states to the 
$(j=5/2)\Gamma_{8}$, 0.672, is large compared with 0.341 which is expected
from the thermal population with their  energy difference, 115K.
Larger HI near $E_{\rm F}$ enhances the population of 
$(j=5/2)\Gamma_{7}$
when the HI of the LDA band is used.

\subsection{ DMFT calculation}

\begin{figure}[htb]
\begin{center}
\includegraphics[width=8cm,clip]{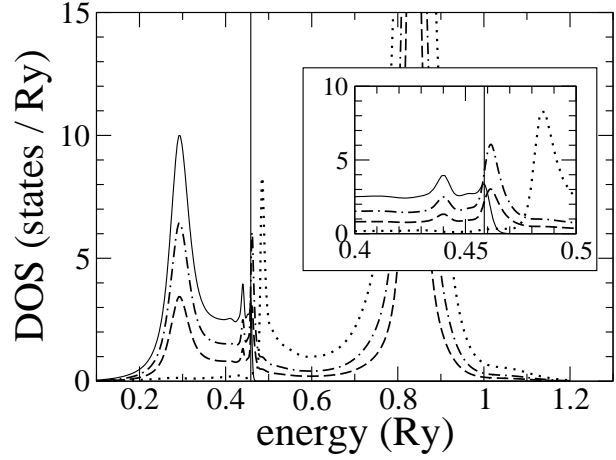}
\end{center}
\caption{$4f$ spectra of the auxiliary impurity problem 
for $\gamma$-Ce.
For the definitions of lines, see the caption of Fig.~\ref{fig:5}.
}
\label{fig:6}
%\end{fullfigure}
\end{figure}

In Fig.~\ref{fig:6},  we show DOS of the $4f$ components 
obtained by the
DMFT band calculation at $T=300$ K.

%Fig.6

%Fig7

\begin{figure}[htb]
\begin{center}
\includegraphics[width=8cm,clip]{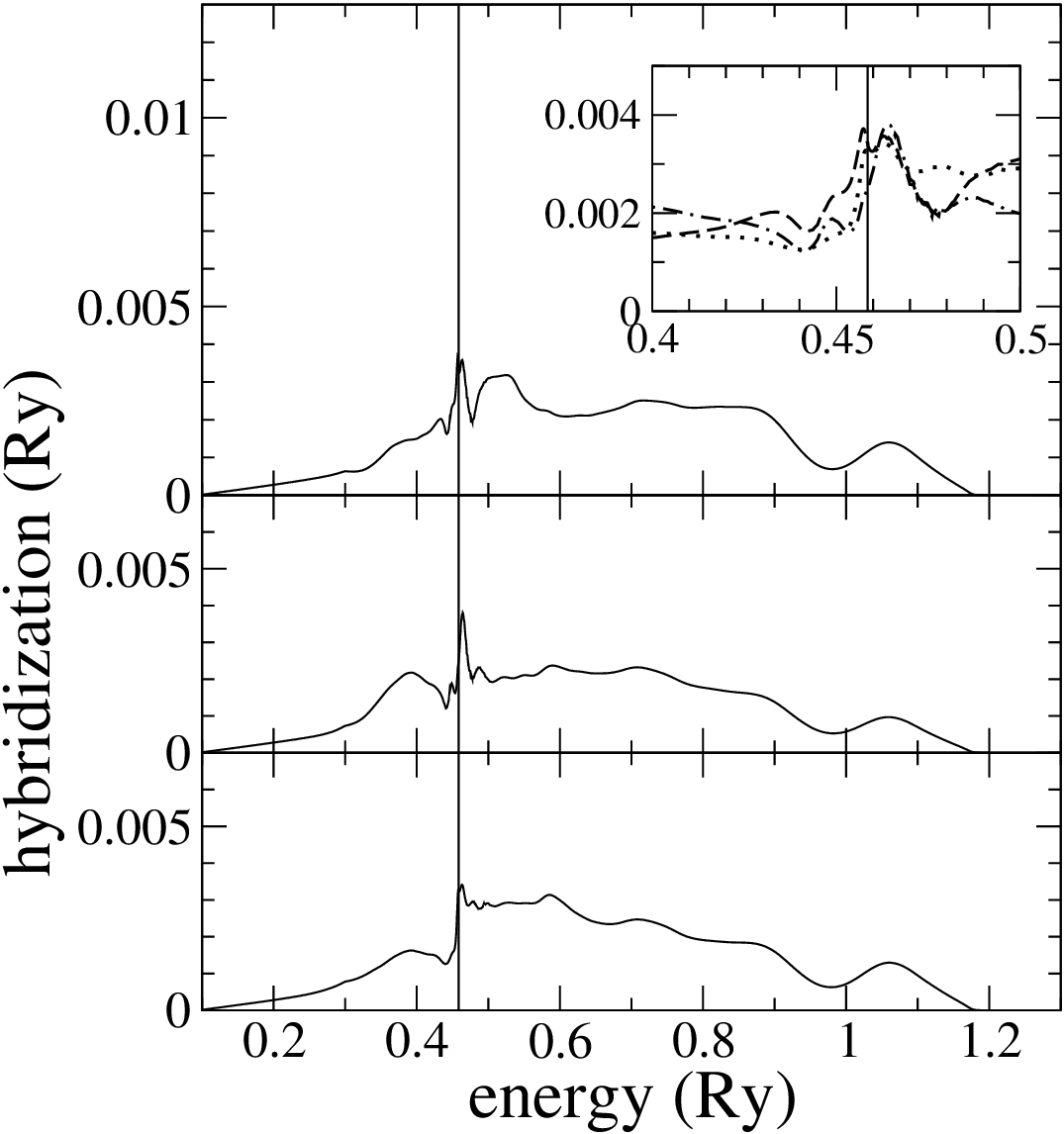}
\end{center}
\caption{Effective HI in DMFT
for $\gamma$-Ce.
For the definitions of lines, see the caption of Fig.~\ref{fig:4}.
}
\label{fig:7}
\end{figure}

The effective HI is shown in Fig.~\ref{fig:7}.
Compared with  the result in Fig.~\ref{fig:4},
the peaks  of  HI at about $0.015$ Ry 
below  $E_{\rm F}$
are reduced.
At the same time the peaks of $(j=5/2)\Gamma_{8}$ and $j=7/2$ components
are pushed up just above $E_{\rm F}$.

The HI in the DMFT has the small peaks (or shoulder for the $j=7/2$ component)
just at $E_{\rm F}$.
The averaged HI near $E_{\rm F}$ is reduced.
Reflecting this 
the peak of the spectra  at $E_{\rm F}$ is relatively 
weak, 
and the SO side peak in PES becomes conspicuous
in the DMFT band.
Such reduction of HI just at $E_{\rm F}$ has been also observed in 
previous calculation~\cite{A20}.

We started calculation by choosing the energy levels of 
$(j=5/2)\Gamma_{7}$ and $(j=5/2)\Gamma_{8}$ to be equal, 
$-0.1500$ Ry. The energy level of $j=7/2$ is initially set 
as $-0.1300$ Ry.
At the first stage, 
(A) HI is fixed as  HI of LDA, and only energy levels are
changed.
We search for the  solution which gives the partial occupation 
numbers of band with the
self energy grossly agree with those of the impurity model.
The target of the total occupation number of $4f$-electron is set 
as $0.982$. 
In the next stage, (B)
the self-consistent spectra and the occupation numbers are 
determined by changing  HI and levels by the iteration method. 
We obtain energy levels in the auxiliary impurity model,
$E((j=5/2)\Gamma_{7})=-0.11943$ Ry,
$E((j=5/2)\Gamma_{8})=-0.12016$ Ry and
$E(j=7/5)=-0.09595$ Ry.
Even when we skip the stage A, we can obtain similar self-consistent 
results, but not exactly identical ones in the strict sense
because the tolerance is not small.
%The iteration process shows complicated behavior in this case 
%because the large change of the partial occupation numbers and the 
%change of HI occur simultaneously.

The occupation numbers in the auxiliary impurity problem and the 
DMFT band are shown in the Table \ref{t1}.
The tolerance for the relative difference of $4f$ DOS is set to be 0.1 for 
the $j=5/2$ group and $1.5\times0.1$ for the $j=7/2$ component.

The energy level of $j=7/2$ states is about 
$0.32$ eV ($= -0.09595-(-0.11943)$ Ry)
higher than the energy  of $j=5/2$.
The SO splitting is slightly enhanced than the value, 0.29 eV  
estimated by the 
SO constant in the LDA band calculation.

We obtained the 
cubic crystalline field splitting in the auxiliary impurity model,
115 K= ( = $-0.11943-(-0.12016))$ Ry,
which  is  caused purely by  
the hybridization effect
because we have not included the electrostatic CF field in the
calculation.
However, this difference should not be considered literally.
The difference in HI will also cause the effective CF energy.
The relative occupation number of $(j=5/2)\Gamma_{7}$
to $(j=5/2)\Gamma_{8}$, $0.3292/0.6165 = 0.53$ is almost equal to the
ratio of their degeneracy factor (0.5), but somewhat larger than it.
This means that their effective energy levels are almost equal, 
but $(j=5/2)\Gamma_{7}$ has slightly lower energy.

%Fig8
\begin{figure}[htb]
\begin{center}
\includegraphics[width=8cm,clip]{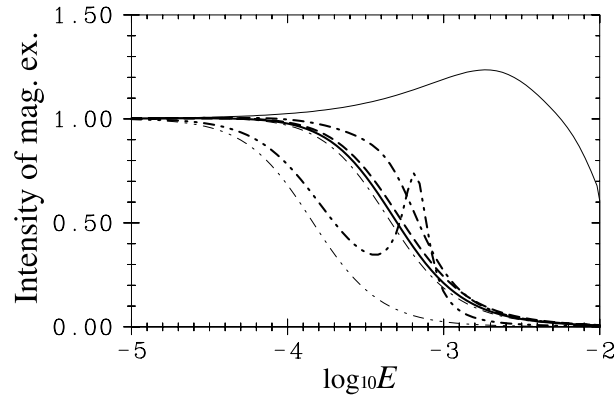}
\end{center}
\caption{
Magnetic excitation spectra calculated by using the
effective HI for various compounds.
The horizontal axis is the logarithm of energy in Ry, and the spectral
intensities 
are normalized by the values at the lowest energy ($E=10^{-7}$ Ry).
The bold (thin) solid line is calculated for $\gamma$-Ce 
($\alpha$-Ce).
The bold  dot-dashed line is calculated for CeSb.
The bold two-dots-dashed is calculated for 
CeSb with reduced Sb $p$ - Ce $d$ band overlapping, see the text.
For these lines, the full matrix elements of magnetic moment 
in the $j=5/2$ manifold are used. 
The bold dashed line is calculated for 
the fictitious model of $\gamma$-Ce in which the matrix 
elements of magnetic moment within the $(j=5/2)\Gamma_{7}$ manifold
are used.
The thin dot-dashed line is calculated for the model of CeSb with 
matrix elements within the $(j=5/2)\Gamma_{8}$ manifold.
The thin two-dots-dashed line is calculated for 
the model of CeSb with the reduced band overlapping and
the matrix elements within the $(j=5/2)\Gamma_{7}$ manifold.
}
\label{fig:8}
\end{figure}

In Fig. \ref{fig:8} we show the magnetic excitation spectra
of the auxiliary impurity model,
$\frac{-1}{\pi}\Im \chi(\omega)\frac{1}{1-\EX^{-\beta\omega}}$,  as a
function of logarithm of energy (in Ry).
The intensity is normalized by the value in the low energy limit.
The bold solid line and the bold dashed line are calculated for the 
effective HI for $\gamma$-Ce at $T=300$K (i.e., Fig. \ref{fig:7}).
The bold solid line is obtained by using the matrix element of the 
magnetic moment within the $j=5/2$ manifold, while the bold dashed line 
is calculated by the fictitious model in which we use the matrix within the 
$\Gamma_{7}$ manifold.
The bold solid line decreases to half of the low energy value 
at $-3.37$, which corresponds to the excitation energy about 67 K.
This energy will be ascribed to the measure of $T_{\rm K}$.
The bold dashed line decreases to half of the low energy value at
slightly higher energy side, -3.32 (76K), but the energy difference is
very small.
When we consider the fictitious model with the matrix elements only
within the $\Gamma_{8}$ states, the magnetic excitation spectra shift to
low energy side by about 10 K.
The CF splitting energy is small compared with $T_{\rm K}$. 

We have also calculated the magnetic excitation 
for the case of Fig. \ref{fig:5}.
It decreases to half of the low energy limit at $-2.6$ which corresponds 
to 400 K.
$T_{\rm K}$ is reduced in the DMFT band.

%As noted previously, 
%HI of $(j=5/2)\Gamma_{7}$ is large just below $E_{\rm F}$,
%but is small in the deeper energy region.
%The CF energy splitting depends on  detailed 
%energy dependence of HI.
%We note the electrostatic crystalline field potential, which is usually 
%considered as the origin of the crystalline field splitting 
%of $4f$ state, is not included in the present calculation.

%\begin{figure}[htb]
%\begin{center}
%%\includegraphics[width=8cm]{flspD.6.eps}
%\includegraphics[width=8cm,clip]{fig/fig7.eps}
%\end{center}
%\caption{$4f$ spectra of the auxiliary impurity problem with  
%effective HI
%for $\gamma$-Ce.
%For the definitions of lines, see the caption of Fig.~\ref{fig:5}.
%}
%\label{fig:7old}
%\end{figure}

The mass renormalization factors and the effective levels
are shown in the Table \ref{t1}.
Here we have approximated as 
$\VEP^{({\rm imp.})}_{\Gamma}
+\Delta\VEP_{\Gamma}
+\tilde{\Sigma}_{\Gamma}(\VEP)-(Z^{-1}_{\Gamma}-1)(\VEP-E_{\rm F}) 
\sim 
\bar{\VEP}_{\Gamma}-(\bar{Z}^{-1}_{\Gamma}-1)(\VEP-E_{\rm F}) $. 
The effective levels are measured from the Fermi energy.
We found DOS by  using these renormalized band 
parameters shows,
only in the restricted energy region 
very near $E_{\rm F}$,
similar value 
to  DOS obtained by the direct DMFT calculation.
The renormalization factor about 40 for $j=5/2$ states is very large.
But we think these values should not be considered literally 
because the temperature $T=300$ K is very high compared with the Kondo 
temperature.
The imaginary part of the self-energy at the Fermi energy is 
estimated to be about 2 eV.
The actual effect will be renormalized as $1/40 \sim 0.05$ eV,
but is still not small.  

The change of the band structure in DMFT will also be important in 
the quantitative calculation of the specific heat.
The
$4f$ component at the Fermi energy in the DMFT
band (about 6 states/Ry, see Fig. \ref{fig:6}) is 
small compared with that (about 35 states/Ry, see
Fig.\ref{fig:3}) 
in LDA.
If one define the mass enhancement factor 
as the ratio to the DOS 
of the LDA calculation, the effective factor will be reduced by 
the quantity,
$\frac{6}{35}$.

At this point we note 
the difficulty in DMFT calculation.
The self-energy of the hybridization is 
relatively  very small,  about 1/100 of the self-energy due to the
Coulomb interaction except the energy region near 
$E_{\rm F}$.
The latter is roughly estimated as 
$Z^{-1}N_{f}v^{2}\rho_{c}$ in the resolvent method, and we have 
$Z^{-1}N_{f} \sim 500$, 
where $Z^{-1}$ is the mass enhancement factor.
This causes trouble in the iteration procedure of DMFT calculation.
Usually the new trial HI is estimated from the 
difference of the  quantities
$G^{-1}_{\rm \Gamma}(\VEP)=\VEP-\VEP_{\Gamma}
-\Sigma^{({\rm C})}_{\rm \Gamma}(\VEP)-
\Sigma^{({\rm h})}_{\rm \Gamma}(\VEP)$
obtained by the band calculation and the impurity problem for the 
trial HI.
(In this place, the self-energy due to the Coulomb interaction is 
denoted as $\Sigma^{({\rm C})}_{\rm \Gamma}(\VEP)$ instead of 
 $\Sigma_{\rm \Gamma}(\VEP)$ to avoid confusion.)
Usually we have relation 
$|\Sigma^{({\rm C})}_{\rm \Gamma}(\VEP)| 
>> |\Sigma^{({\rm h})}_{\rm \Gamma}(\VEP)|$, 
then the very small uncertainty of 
$\Sigma^{({\rm C})}_{\rm \Gamma}(\VEP)$ greatly affects  new trial of 
$\Sigma^{({\rm h})}_{\rm \Gamma}(\VEP)$, 
while
the self-energy 
$\Sigma^{({\rm C})}_{\rm \Gamma}(\VEP)$ itself is very sensitive to 
the detail of 
HI.
These facts lead to the unstable iteration loop.
The change of the new trial HI from that of the preceding 
step must be restricted
to be very small.
Therefore, the convergence to small tolerance is very slow.
The fact 
$|\Sigma^{({\rm C})}_{\rm \Gamma}(\VEP)| 
>> |\Sigma^{({\rm h})}_{\rm \Gamma}(\VEP)|$
means, in other words,  that any trial HI usually gives 
the almost self-consistent solution
except very near $E_{\rm F}$.

We  do not allow  the large change of HI except for the 
 energy region near $E_{\rm F}$~\cite{A58}.
We also stopped the  DMFT iteration steps less than 60 though the
maximum relative difference of the spectral density is not so small, 
about 0.1  after the smoothening process of spectra~\cite{A59}.
Usually difference appears in the  steeply increasing 
(or decreasing) region of the sharp peaks of the spectral density.
The difference  can not be reduced easily by further iteration steps.
However, the difference of the peak height itself is usually 
small.
In addition, the relative difference of the densities just 
at $E_{\rm F}$ usually becomes less than 0.01.
The tolerance for the $j=7/2$ component is set about 3 times larger
because maximum difference of it appears near the SO side peak  
in the  unoccupied energy region not near
 $E_{\rm F}$.

To give the atomic $4f$ electron number to be near unity (0.983),
we obtained the $4f$ level to be about $-0.120$ Ry below the Fermi energy.
The calculated $4f$ peak locates  at about 2.2 eV below 
$E_{\rm F}$,
which is slightly deep  compared with the energy 2 eV
observed in PES experiments~\cite{B1,A23,A24}.
If we need the $4f$ level to appear near the experimental position,
the $4f$ occupation number must be chosen to be $0.97$.
%The occupation number of $4f$ electrons becomes about ****, which is very
%small compared with the value obtained by LDA calculation.
%We think  that the total occupation of the $4f$ electron should not so
%largely change from the value of LDA, because large electrostatic energy
%had been already included in the LDA band.

%Fig9

%\begin{figure}[htb]
%\begin{center}
%%\includegraphics[width=8cm]{flsp.080.eps}
%\includegraphics[width=8cm,clip]{fig/fig9.eps}
%\end{center}
%\caption{$4f$ spectra calculated by reduced HI ($r=0.7$)
%for $\gamma$-Ce.
%For the definition of $r$, see the text.
%For the definitions of lines, see the caption of Fig.~\ref{fig:5}.
%}
%\label{fig:9}
%\end{figure}

In preceding DMFT calculations, a hump  at $E_{\rm F}$
is obtained by QMC calculation~\cite{A4}, 
but the sharp structure has been 
smeared out because of the broadening and  higher temperature 
calculations.
The Kondo peak  has not been observed in NCA calculation.
In experimental studies, SO side peak  similar to the 
result in Fig.~\ref{fig:7} is observed in the high resolution 
experiments~\cite{B1,A23,A24}.
In the present calculation we have obtained that the CF splitting energy 
is smaller than $T_{\rm K}$.
But this result should be checked carefully because we have neglected 
the electrostatic energy part of the CF splitting.

Detailed comparison with experiments will be carried out in the near
future.

\section{$\alpha$-Ce
}

\begin{figure}[htb]
\begin{center}
\includegraphics[width=8cm,clip]{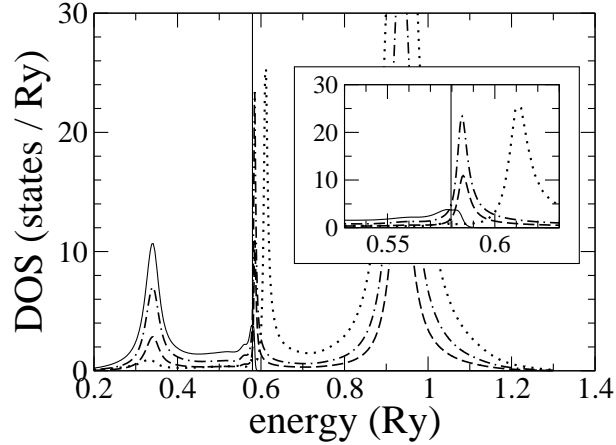}
\end{center}
\caption{$4f$ spectra of the auxiliary impurity model
for $\alpha$-Ce.
For the definitions of lines, see the caption of Fig.~\ref{fig:5}.
}
\label{fig:9}
\end{figure}

Next we study  the  Ce metal with the lattice constant 
of $\alpha$ phase ($a=9.16517$ AU).
The total occupation number of the $4f$ electron in the LDA is $1.053$
which is almost equal to that of $\gamma$-Ce, but is  0.033 smaller.
We choose the target of the $4f$ electron number as $0.95=0.983-0.033$
in $\alpha$-Ce. 
The obtained  occupation numbers and the energy levels 
are summarized in Table \ref{t2}.
The $4f$ DOS is shown in Fig. \ref{fig:9}.
The small SO side peak is identified in PES.
This result is similar to the result in experiments,
but the position of the $f^{0}$ excitation peak is very deep compared 
with  experimental results.  

%Fig10

%Table2

\begin{table}[t]
\caption{Various quantities obtained in DMFT calculation for $\alpha$-Ce.
$E_{\rm F}=0.5796$ Ry, and energy levels are given in Ry.
$\VEP^{\rm a}_{4f}=-0.16200$ Ry.
$\zeta_{4f}=0.007058$ Ry.
For the definition of 
quantities, see the caption of Table I  
}
\label{t2}
\begin{halftabular}{@{\hspace{\tabcolsep}\extracolsep{\fill}}cccc} \hline
%\multicolumn{4}{c} {$r=1$} &\multicolumn{3}{c} {$ r=0.7$} \\
& $\Gamma_{7}$ & $\Gamma_{8}$ & $j=7/2$ 
\\ \hline
$n_{\Gamma}^{\rm LDA}$      &
0.2260   & 0.3725   & 0.4545 
                             \\
$n_{\Gamma}^{({\rm imp'.})}$  &
0.2345   & 0.4132   & 0.2670
                               \\
$n_{\Gamma}^{({\rm band})}$ & 
0.2372   & 0.5654   & 0.1478
                               \\
$n_{\Gamma}^{({\rm imp.})}$ & 
 0.2366  & 0.5658   & 0.1480
                               \\
$\VEP_{\Gamma}$                &
-0.15868  &-0.16027   &-0.13317
                               \\  
$\bar{Z}_{\Gamma}^{-1}$        & 
12.6     & 13.5  & 2.0
                               \\
$\bar{\VEP}_{\Gamma}$            &
0.10320  &0.09076 & 0.24327
                               \\
$\bar{\Gamma}_{\Gamma}$           &
0.0176  &0.0203 & 0.0004
                               \\ 
\hline
\end{halftabular}
\end{table}

%The CF splitting in the auxiliary impurity model is calculated as 
%$180$ K ($= -0.21327-(-0.214431))$ Ry.
%The obtained spectra is shown in Fig.~\ref{fig:10}.
%The peak at the Fermi energy is larger than the peak of 
%the SO side band.
%If we used the HI of LDA band, the SO side band merges to the 
%peak at the Fermi energy.

The magnetic excitation spectra 
are 
shown by the thin solid  line 
in Fig. \ref{fig:8}, 
which show peak at $-2.7$ (310 K) and decrease
at $-1.98$ (1650 K) to half of the low energy limit.
When  we use the fictitious model with 
matrix elements within the $\Gamma_{8}$ manifold, 
the spectrum intensity of the peak region is reduced somewhat.

%Fig11

\begin{figure}[htb]
\begin{center}
\includegraphics[width=8cm,clip]{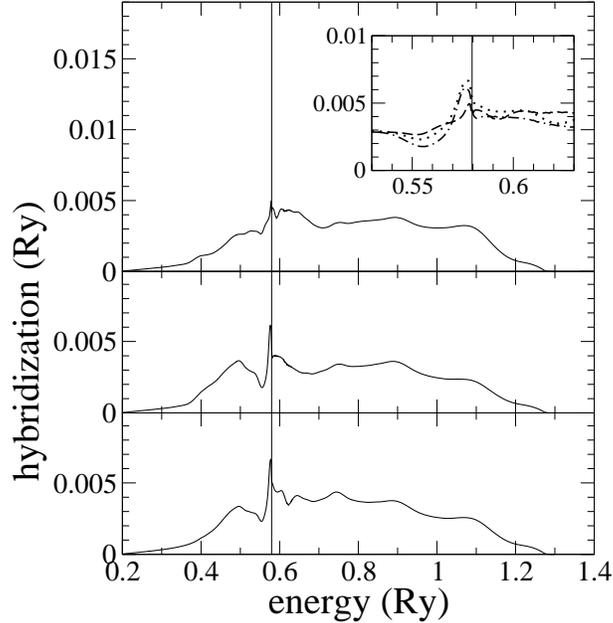}
\end{center}
\caption{Effective HI in DMFT for $\alpha$-Ce.
For the definitions of lines, see the caption of Fig.~\ref{fig:4}.
}
\label{fig:10}
\end{figure}

%Fig12

%\begin{figure}[htb]
%\begin{center}
%%\includegraphics[width=8cm]{flspA7.4.eps}
%\includegraphics[width=8cm,clip]{fig/fig12.eps}
%\end{center}
%\caption{$4f$ spectra calculated by reduced HI($r=0.7$) 
%for $\alpha$-Ce.
%For the definition of $r$, see the text.
%For the definitions of lines, see the caption of Fig.~\ref{fig:5}.
%}

%\label{fig:12}
%\end{figure}

In Fig.~\ref{fig:10}, we show the effective HI.
Peaks of HI near $E_{\rm F}$ in LDA are reduced to the narrow peaks just
on $E_{\rm F}$ in the DMFT band.
Thus $T_{\rm K}$ becomes lower than that obtained by using HI of LDA band.

The position of the $f^{0}$ peak appeared at about 3.2 eV 
below $E_{\rm F}$.
This is deep compared with the experimental result of about 2.4 eV.
To obtain the $f^{0}$ level at about 2.7 eV,
the occupation number of $4f$ electron must be reduced to 0.8.
This is very
small compared with the value obtained by LDA calculation.
We think  that the total occupation of the $4f$ electron should not so
largely change from the value of LDA, because large electrostatic energy
had been already included in the LDA band.
The HI calculated directly from the band calculation seems to be somewhat
large.

\section{CeSb
}
CeSb has NaCl structure in which Ce ions form the {\it fcc} lattice.
The valence  and the conduction bands mainly consist of the 
Sb $5p$ states and the Ce $5d$ states, respectively.
They overlap slightly and CeSb has semi-metallic band structures.
The top of the $p$ valence band has the $(j=3/2)\Gamma_{8}$ character, 
and thus has the stronger hybridization matrix with the $\Gamma_{8}$ CF 
state.

In the $p-f$ mixing model, 
a part of the $\Gamma_{8}$ valence 
band which has the same symmetry with the occupied $4f$ band
is pushed up above 
$E_{\rm F}$
 due to the hybridization
in the ordered phase\cite{A30,A31,A32,A33}.
Such re-constructed electronic band structures in the ordered states 
are expected by Hatree-Fock-like static mean field theory~\cite{A45}, 
and confirmed by experimental studies~\cite{A34,A36,A40}.
We need the DMFT band calculation in the paramagnetic phase.

%Fig13

\begin{figure}[htb]
\begin{center}
\includegraphics[width=8cm,clip]{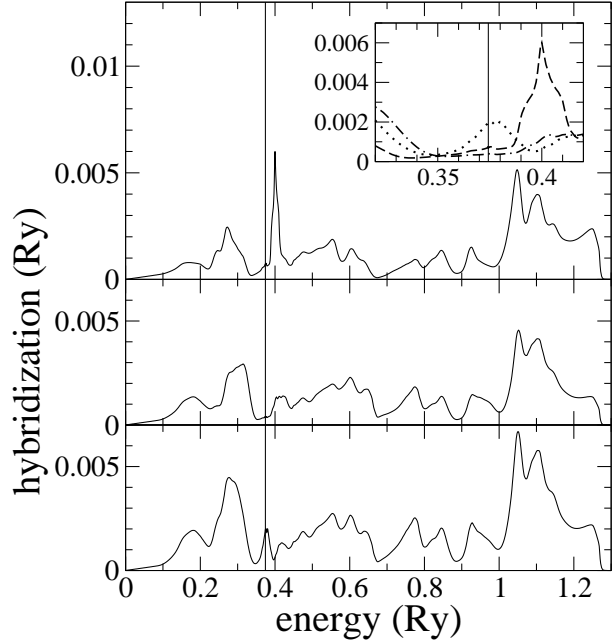}
\end{center}
\caption{The HI calculated by the LDA band for CeSb.
For the definitions of lines, see the caption of Fig.~\ref{fig:4}.
}
\label{fig:11}
\end{figure}

%Fig14

\begin{figure}[htb]
\begin{center}
\includegraphics[width=8cm,clip]{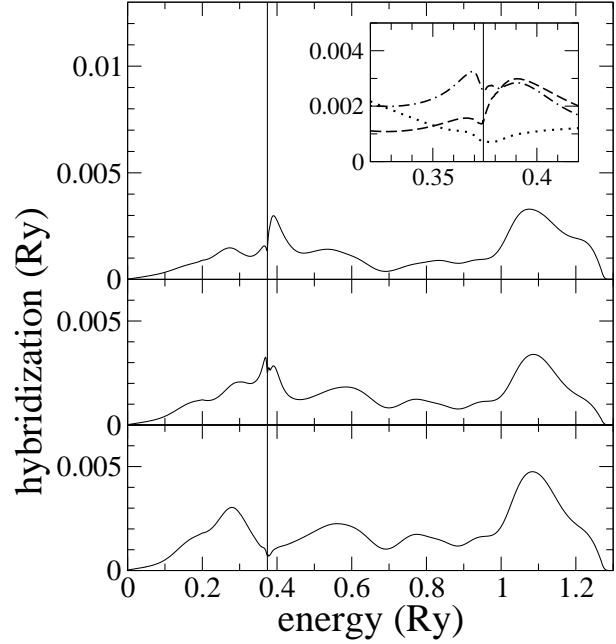}
\end{center}
\caption{Effective hybridization in DMFT for CeSb.
For the definitions of lines, see the caption of Fig.~\ref{fig:4}.
}
\label{fig:12}
\end{figure}

\begin{figure}[htb]
\begin{center}
\includegraphics[width=8cm,clip]{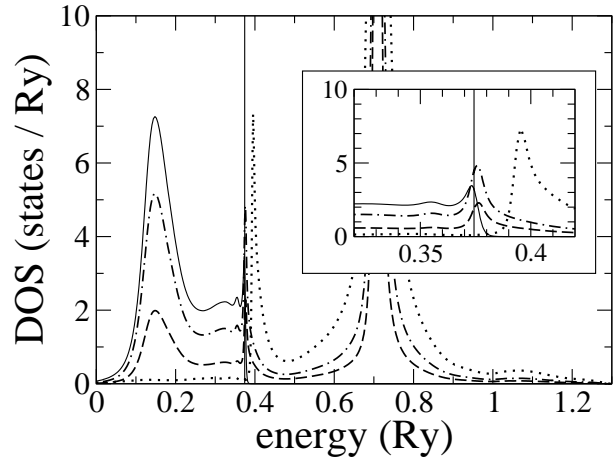}
\end{center}
\caption{$4f$ spectra of the auxiliary impurity model
for CeSb.
For the definitions of lines, see the caption of Fig.~\ref{fig:5}.
}
\label{fig:13}
\end{figure}

In Fig. \ref{fig:11}, we show
the HI for  CeSb of LDA band, and 
in Fig. \ref{fig:12} the effective HI
of DMFT calculation.
The peak of HI in the energy region $0.25 \sim 0.3$ Ry  in Figs. \ref{fig:11} 
and \ref{fig:13} mainly 
corresponds to the  $j=3/2$ component of Sb $p$, and the peak at about 
$0.15$ Ry corresponds to the $ j=1/2$ component~\cite{A27,A26}.

In Fig. \ref{fig:11}, we have  peak structures near $E_{\rm F}$
which are caused by the hybridization of band states with 
the $4f$ bands.
The
$4f$ bands are pinned near  $E_{\rm F}$ in the paramagnetic state.
But the peaks do not necessarily appear just at $E_{\rm F}$ in the 
LDA band.
We have sharp peaks just at $E_{\rm F}$ in the DMFT calculation as 
shown in Fig.~\ref{fig:12}.
This may be caused by the hybridization with the correlated $4f$ band.

The occupation number ratio of the $(j=5/2)\Gamma_{7}$ to the 
$(j=5/2)\Gamma_{8}$, 0.39 is smaller than ratio 1/2 of their 
degeneracy factor, thus the effective energy of the latter is lower 
in this case.

%Fig15

In the single particle spectra shown by Fig. \ref{fig:13},
a small hump is observed at about 0.7 eV below $E_{\rm F}$.
A large peak appears at about 3 eV below $E_{\rm F}$.
This double-peak-like structure is qualitatively similar to 
the result of experiments in  PES.
The peak structure at $E_{\rm F}$ in HI of DMFT 
causes the sharp peak just at $E_{\rm F}$ in the spectra.

We show the magnetic excitation spectra by the bold dot-dashed line
in Fig. \ref{fig:8}.
The excitation spectra for the fictitious model with matrix elements
within the $(j=5/2)\Gamma_{8}$ manifold are shown by the thin dot-dashed
line.
The later decreases to the half of the low energy value at the
excitation energy $-3.4$ (63 K), while the former has contribution 
from the $(j=5/2)\Gamma_{8} \rightarrow (j=5/2)\Gamma_{7}$ excitation 
in the energy region about $-3.3$ (79 K).

%Fig16

\begin{figure}[htb]
\begin{center}
\includegraphics[width=8cm,clip]{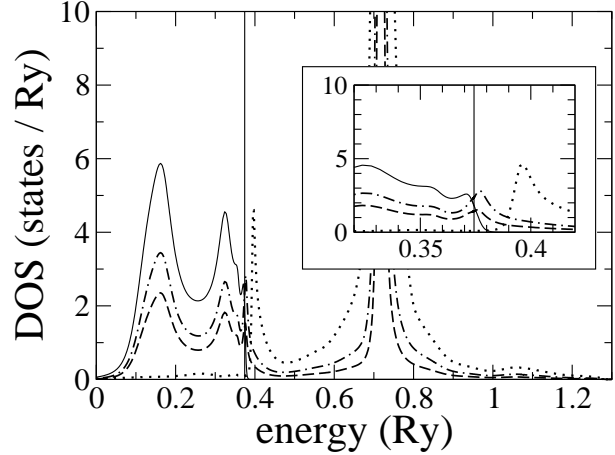}
\end{center}
\caption{$4f$ spectra calculated by reducing the band overlapping 
between Sb $p$ and Ce $d$ for CeSb.
The Sb $p$ and $s$ bands are shifted to low energy side 0.5 eV by putting 
$\omega_{Sb,s}(-)=\omega_{Sb,s}(-)=-0.037$ Ry.
The atomic energy level of $4f(j=5/2)\Gamma_{7}$ is shifted $-1000$ K.
For the definitions of lines, see the caption of Fig.~\ref{fig:5}.
}
\label{fig:14}
\end{figure}

%Table3

\begin{table}[t]
\caption{Various quantities obtained in DMFT calculation for CeSb.
$E_{\rm F}=0.3743$ Ry, and energy levels are given in Ry.
$\VEP^{\rm a}_{4f}=-0.16022$ Ry.
$\zeta_{4f}=0.007043$ Ry.
For the definition of 
quantities, see the caption of Table I  
}
\label{t3}
\begin{halftabular}{@{\hspace{\tabcolsep}\extracolsep{\fill}}cccc} \hline
%\multicolumn{4}{c} {$r=1$} &\multicolumn{3}{c} {$ r=0.7$} \\
&$\Gamma_{7}$ & $\Gamma_{8}$ & $j=7/2$ 
\\ \hline
$n_{\Gamma}^{\rm LDA}$      &
0.3788   & 0.4994   & 0.3178
                            \\
$n_{\Gamma}^{({\rm imp'.})}$  &
0.2753   & 0.6917   & 0.0458
                               \\
$n_{\Gamma}^{({\rm band})}$ & 
0.2652   & 0.6892   & 0.0421
                              \\
$n_{\Gamma}^{({\rm imp.})}$ & 
 0.2658  & 0.6900   & 0.00407
                              \\
$\VEP_{\Gamma}$                &
-0.15815  &-0.15827   &-0.113914
                               \\  
$\bar{Z}^{-1}_{\Gamma}$        & 
23.1     & 29.9 & 2.9 
                               \\
$\bar{\VEP}_{\Gamma}$            &
0.17443  &0.13428 & 0.28134
                               \\  
$\bar{\Gamma}_{\Gamma}$            &
0.1506  &0.1892 & 0.0006
                               \\ 
\hline
\end{halftabular}
\end{table}

The mass enhancement factors are estimated to be very large.
However, these large values will not have actual 
meaning as already noted 
because we have calculated them at the high temperature region, $T=300$ K.

At this point we should note that the present calculation for 
CeSb has started
from the LDA band.
Usually it leads to rather large DOS near $E_{\rm F}$ due to
the large band overlapping of Sb $p$ and Ce $d$ bands~\cite{A45,D3,D5}.
Very high $T_{\rm K}$  will be partly ascribed to the use of this 
large DOS.
In transport experiments, the Kondo temperature is expected to be 
about 10 K~\cite{A32}.
The correct estimation of the $T_{\rm K}$ will depend on this 
band overlapping.

We tentatively shift the Sb $5s$ and $5p$ bands $-0.5$ eV
by putting $\omega_{{\rm Sb},p}(-)=\omega_{{\rm Sb},s}(-)=-0.037$ Ry~
\cite{D5}.
The atomic energy level of $4f(j=5/2)\Gamma_{7}$ is also 
shifted $-1000$ K.
This extra shift  may correspond to the CF energy splitting due to the
electrostatic potential.
The DMFT calculation for this modified model is carried out.
The  occupation numbers of $(j=5/2)\Gamma_{7}$ and $(j=5/2)\Gamma_{8}$ 
are obtained as  0.39 and 0.58, respectively.
The effective CF energy becomes lower for 
$(j=5/2)\Gamma_{7}$ in this case.
We note, however, that we obtain  another type solution in which
the occupied
electrons are almost $(j=5/2)\Gamma_{8}$ when the extra CF shift 
of $(j=5/2)\Gamma_{7}$ is reduced to $-500$ K.
Both types of states are competing.

The single particle spectra are shown in Fig. \ref{fig:14}.
They show the shallow energy peak at about 0.8 eV and the deep energy
peak at about 3 eV below $E_{\rm F}$.  
This result well corresponds to the experimental result.
The magnetic excitation is shown in Fig. \ref{fig:8}
by the bold two-dots-dashed line.
The excitation of the fictitious model with the matrix elements 
within the 
$(j=5/2)\Gamma_{7}$ manifold is shown by the thin two-dots-dashed line.
The later decreases to the half of the low energy value at about 
excitation energy $-3.9$ (20K), while the former has a peak at 
the excitation energy about $-3.2$ (100K) corresponding to the 
$(j=5/2)\Gamma_{7} \rightarrow (j=5/2)\Gamma_{8}$ transition.
We note that the peak of HI at $E_{\rm F}$
is considerably small and sharp  compared with the result in
 Fig. \ref{fig:12}.
It becomes  very similar to 
the peak in HI used by Takeshige to reproduce 
$T_{\rm K}$ of CeSb~\cite{A54}.

Detailed systematic studies on PES and Fermi surface structures 
are necessary in quantitative calculation.
However, we note again, the hybridization with the correlated $4f$ 
band will produce sharp peaks at $E_{\rm F}$ in HI. 
The similar result has been also found in ref. \citen{D2}.
The such effect has not been considered in previous studies~\cite{A27}.
It is also necessary to analyze the transport phenomena 
by including the contribution of the correlated $4f$ band.

\section{Summary and Discussion}

We have developed a band calculation method based on the 
DMFT + LMTO frame work, 
and applied it to Ce metal and CeSb.
The starting band is calculated by the LDA method,
 and the auxiliary  impurity problem is solved 
by the NCA$f^{2}$v' method 
which is explained in  Appendix.
The method can include the $f^{2}$ configuration and gives
the  Kondo temperature ($T_{\rm K}$) 
not the so different from that obtained by NRG, thus 
 we can expect quantitative 
calculation.
The spin-orbit (SO) splitting and the crystalline field (CF) splitting
are included in the self-energy term.
$T_{\rm K}$ is estimated from the calculation of the magnetic
excitation spectra.

In Ce metal, 
the total hybridization intensity (HI) near $E_{\rm F}$
is reduced in DMFT.
Reflecting this the SO side peak in PES becomes relatively conspicuous
compared to the calculation directly using HI
 in the LDA band. 
The difference of HI between $(j=5/2)\Gamma_{7}$ and 
$(j=5/2)\Gamma_{8}$ states leads to the energy level difference between them
in the auxiliary impurity model.
But in magnetic excitation, this difference does not directly 
appear because the HI itself causes the CF splitting.  
The Kondo temperature for $\gamma$-Ce is estimated to be about 70 K, and
the CF splitting is smaller compared with this in the present
calculation.
This point should be checked carefully, however, because we have not
considered the CF splitting energy due to the electrostatic potential.  
The $T_{\rm K}$ in $\alpha$-Ce is estimated to be about 1700K.
In experiments the magnetic excitation in $\alpha$ Ce is expected to be 
about $10^{3}$ K~\cite{E1}.
The SO splitting in DMFT calculation for Ce metal is larger than 
$T_{\rm K}$, thus the
Kondo temperature  is greatly reduced 
from that estimated by calculation neglecting the SO 
effect.
The studies of $\alpha$-$\gamma$ transition including the effect of SO 
and CF splittings are retained to the future.

For CeSb, we have obtained similar results to results of the previous 
calculation based on the impurity model in the gross sense.
The double-peak-like structure in  PES experiments is reproduced.
We have a peak  very near $E_{\rm F}$ in HI.
It leads to the higher $T_{\rm K}$.
The  peak is caused by the hybridization with the 
correlated $4f$ band in DMFT theory.

When the band overlapping between the Sb $p$ and Ce $d$ is reduced, 
the double peak structure of PES and
$T_{\rm K}$ about 20 K were obtained. 
In transport experiments, the Kondo temperature is expected to be 
about 10 K~\cite{A31}.
Systematic studies of band structure such as analysis of the dHvA 
effect~\cite{A34,A36}, 
the angle-resolved-PES~\cite{D1} and transport properties~\cite{A31} 
will be carried out in the
near future.

In both materials,
the HI which is directly calculated by the LDA band calculation seems to be 
larger than the magnitude expected from the comparison with 
PES experiment.
But in the DMFT calculation the effect of HI is relatively 
weakened.
The calculated HI is somewhat lager, but not so drastically large 
than the magnitude expected from experimental data.

The auxiliary impurity problem is solved by the NCA$f^{2}$v' method.
It gives not so different excitation spectra from those calculated by
NRG.
However the NCA$f^{2}$v gives the improved result as shown in
 Fig.~\ref{fig:2}.
The implementation of this method will also be desirable though it 
will need huge computation time.

\section*{Acknowledgments}

The authors would like to thank to Professor H. Shiba,
Professor J. Flouquet,
Professor T. Fujiwara and Dr. Ferdi Aryasetiawan
for stimulation,
and to Professor Y. Kuramoto for important  comments on the resolvent
method.
The numerical computation was partly performed in the Computer Center
of Tohoku University and the Supercomputer Center of Institute for Solid
 State Physics (University of Tokyo).
This work was partly supported by JSPS and MEXT, KAKENHI(No.14340108), 
(No. 14540322)
and (No.15034213), and the selective research fund of Tokyo Metropolitan 
University.

\vfill\eject

\appendix
\section{NCA$f^{2}$v' method
}
Let us define the Hubbard X operator
between the atomic states $a$ and $b$,
\Beqa
X(a,b)=| a >< b |.
\Eeqa
We denote   by  $| 0 >$
the $f^{0}$ state,
and by $|\eta >$
the states 
in which the $\eta$  orbital is occupied. 
When the two electrons occupy the $\eta$ and $\eta'$ orbitals, 
the state is denoted as $|\eta\eta' >$.
The atomic state of the $4f$ electron is expressed as 
\Beqa
H_{\rm f} =\sum_{a} E(a) X(a,a),
\Eeqa 
where $E(a)$ is the energy of the atomic state.
The creation operator of the $f$ electron is also
expressed as 
\Beqa
f^{+}_{\eta}=<\eta | f^{+}_{\eta} | 0 > X(\eta, 0) 
+\sum_{\eta'(\neq\eta)}
<\eta\eta'|f^{+}_{\eta} |\eta' > X((\eta\eta'), \eta').
\Eeqa

%Fig.A.1

\begin{figure}[htb]
\begin{center}
\includegraphics[width=8cm,clip]{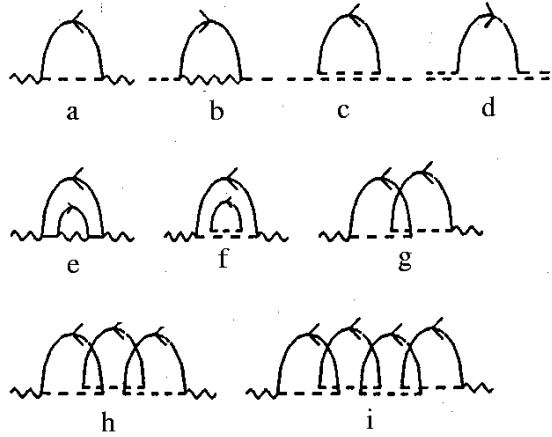}
\end{center}
 \caption{Typical diagrams in the resolvent method.
top: the second order diagram for the resolvent self-energies.
The wavy line denotes the $f^{0}$ resolvent, the dashed line 
and the double dashed line denote  the $f^{1}$ and 
the $f^{2}$ resolvents, respectively.
middle: the fourth order diagrams for the self-energy of 
$f^{0}$ resolvent.
bottom: higher order diagrams for the $f^{0}$
resolvent, which lead to the vertex correction.
}
\label{fig:15}
\end{figure}

\begin{figure}[htb]
\begin{center}
\includegraphics[width=6cm,clip]{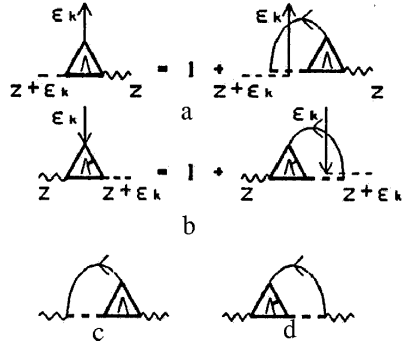}
\end{center}
 \caption{Graphical representations of the integral equation 
for the vertex function (top and middle panel),
and the equation for the resolvent self-energy of $f^{0}$ (bottom).
}
\label{fig:16}
\end{figure}

The resolvent for the atomic state 
$a$, $R(a ;z)$
 is expressed by using the self energy of the resolvent 
$\Sigma(a;z)$ as~\cite{C7,C10,A16}
\Beqa
R(a;z)=\frac{1}{z-E(a)-\Sigma(a;z)}.
\label{eq.A.4}
\Eeqa
The skeleton diagrams of the NCA approximation which include the 
$f^{2}$ configuration are given in the top panel of Fig.
\ref{fig:15}~\cite{C10}. 
The original NCA for $f^{0}-f^{1}$ fluctuation model is 
given by the 
diagram a and b.
In the diagram a for the self-energy of the $f^{0}$ state,
the fermion loop which
consists of the solid and the dashed lines appears. 
It leads to the summation on the degeneracy factor, 
and has the large factor $N_{f}$.
Based on this fact, the self-consistency solutions of 
a and b types are derived.
A part of the fourth order terms for the 
$f^{0}$ self-energy is given in the middle panel of Fig.
\ref{fig:15}.
The diagrams a and b can be included by the skeleton diagram a of the 
top panel.
 However, the diagram c cannot be included in it.
This diagram has also two loops, so it will have $N_{f}^{2}$ same as
the a and b have.
In the higher order diagrams, we must include the diagrams shown in the 
bottom panel of Fig. \ref{fig:16}.
Diagrams of this type for the $f^{0}$ self-energy
can be summed up when we solve the integral equation 
for the vertex part shown graphically by top or middle  panels of
Fig. \ref{fig:16}~\cite{A16,F1},
\Beqa
\Lambda(\eta ; z, x) = 
1+ \sum_{\eta'(\neq\eta)}\int {\rm d}x' M_{\eta'}(x')
R(\eta\eta' ; z+x+x' ) R(\eta' ; z+x')f(x') 
\Lambda(\eta'; z, x' ),
\Eeqa
where we have used the type a equation in the Fig. \ref{fig:16}.
The quantity
$M_{\eta}(x)$ denotes the hybridization intensity
$v^{2}\rho_{c}(x)$.

%Fig.A.2

The graphical expression of the $f^{0}$ self-energy is shown in the 
bottom panel in Fig. \ref{fig:16}, and the equation is given as~\cite{A16}, 
\Beqa
\Sigma(0 ; z )= \sum_{\eta}\int {\rm d}x M_{\eta}(x)
R(\eta ; z+x)f(x)\Lambda(\eta ; z, x ).
\label{eq.A.6}
\Eeqa
When one put $\Lambda(\eta ; z, x) = 1$ by neglecting the 
contribution from the $f^{2}$ state, eq. (\ref{eq.A.6} ) corresponds to the 
equation in the usual NCA of $(f^{0}-f^{1})$ fluctuation.

The self-energies of the $|\eta>$ state and $|\eta\eta'>$ are 
given~\cite{A16} 
\Beqa 
\Sigma(\eta ; z )
=\sum_{\eta'(\neq\eta)} \int {\rm d}x M_{\eta'}(x)
R(\eta\eta' ; z+x )f(x)
\nonumber \\ 
+\int {\rm d}x \Lambda(\eta ; z-x, x)R(0 ; z-x)
 \Lambda(\eta ; z-x, x)f(-x),
\label{eq.A.7} 
\\
\Sigma(\eta\eta' ; z) 
=\int {\rm d}x( M_{\eta}(x)R(\eta'; z-x )
 + v^{2}_{\eta'}R(\eta ; z-x))f(-x)
\nonumber \\
 +\int {\rm d}x {\rm d}x'
   M_{\eta}(x)M_{\eta'}(x')R(\eta' ; z-x')\Lambda(\eta ; z-x-x', x)
\nonumber \\
 \times R(0 ; z-x-x' )\Lambda(\eta' ; z-x-x', x')R(\eta ; z-x)
f(-x)f(-x').
\label{eq.a.8}
\Eeqa

%Fig.A.3
\begin{figure}[htb]
\begin{center}
\includegraphics[width=6cm,clip]{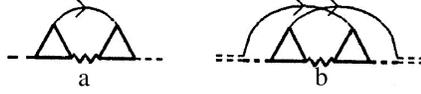}
\end{center}
\caption{Graphical representation for the resolvent 
self-energies of $f^{1}$(left panel) and $f^{2}$(right panel) states.
}
\label{fig:17}
\end{figure}

The graphical expression is shown in Fig. \ref{fig:17}.
In ref. \citen{A16}, it has been  pointed out that
the exchange coupling due to the virtual excitation to the
$f^{2}$ can be led 
even when the $x$ dependence of 
$\Lambda(\eta ; z, x)$ is neglected.
The equation for the vertex is simplified as 
\Beqa
\Lambda(\eta ; z) = 
1+ R(2)^{*}\sum_{\eta'(\neq\eta)}\int {\rm d}x' M_{\eta'}(x')
 R(\eta' ; z+x')f(x') 
\Lambda(\eta'; z ).
\label{eq.A.9}
\Eeqa
where $R(2)^{*}$
is some representative value of the resolvent of the $f^{2}$ state 
for $z \sim \VEP_{f} $.
We usually put as 
$R(2)^{*} = 1/(\VEP_{f}-(2\VEP_{f}+U))$.
The equations (\ref{eq.A.6}) $\sim$ (\ref{eq.A.9})
and (\ref{eq.A.4}) are solved.
As discussed in ref. \citen{A16}, this scheme gives the exchange
constant consistent with that of the S-W transformation in the 
leading order of the $1/N_{f}$ expansion.
We note that the original NCA is correct to the next order of the 
$1/N_{f}$ expansion if the valence fluctuation is restricted within 
$f^{1} - f^{0}$.

The single particle excitation spectrum is given by calculating 
Green's function defined by 
\Beqa
G_{\eta}=-< T_{\tau}f_{\eta}(\tau)f^{+}_{\eta} >.
\Eeqa
By substituting eq. (A.3) into this expression, 
it is decomposed to Green's function defined as 
\Beqa
G(ab,b'a';\tau)=-<T_{\tau}X(ab;\tau)X(b'a')>.
\Eeqa
The diagonal part of the spectral function for the $\eta $ particle 
excitation is given as the convolution integral
\Beqa
\Im G(0\eta,\eta 0: x+{\rm i }0)
=\frac{1+\EX^{-\beta x}}{\pi}
\int {\rm d} y \xi(0; y)(\Im R(\eta ; y+x+{\rm i}0)),
\\
\Im G(\eta'(\eta\eta'),(\eta\eta')\eta': x+{\rm i }0)
=\frac{1+\EX^{-\beta x}}{\pi}
\int {\rm d} y \xi(\eta'; y)(\Im R(\eta\eta' ; y+x+{\rm i}0)),
\Eeqa
where the quantities 
$\xi(a ; x)$ is called as the defect part defined as 
\Beqa
\xi(a ; x) = 
\frac{1}{Z_{f}}\EX^{-\beta x}(-\frac{1}{\pi}
\Im R(a ; x+{\rm i}0)).
\Eeqa

%Fig.A.4

\begin{figure}[htb]
\begin{center}
\includegraphics[width=6cm,clip]{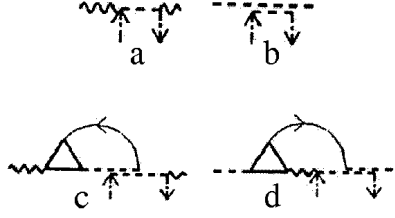}
\end{center}
\caption{Graphical representation of 
the single particle excitation.
The top panel is the NCA$f^{2}$v' part, and 
the bottom is off-diagonal part which are 
considered in NCA$f^{2}$v method.
}
\label{fig:18}
\end{figure}
   
The graphical expression is given by the top panel in the Fig. \ref{fig:18}.
We note that the spectrum intensity has the off-diagonal type term
shown by the bottom in the Fig. \ref{fig:18}, 
which is given as 
\Beqa
\Im G(\eta'(\eta\eta'), \eta 0; x+{\rm i}0)
=\Im G(0\eta,(\eta\eta')\eta'; x+{\rm i}0) \nonumber \\
=-\frac{1+\EX^{-\beta x}}{\pi Z_{f}}\int {\rm d}x'
M_{\eta'}(x')
f(-x')
\int {\rm d}y \EX^{-\beta y}
\nonumber \\
\times \Im(R(\eta'; y+{\rm i}0)
       \Lambda(\eta' ; y-x'+{\rm i}0)R(0; y-x'+{\rm i}0)) \nonumber \\
\times      \Im(R(\eta; y+x-x'+{\rm i}0)R(\eta\eta'; y+x+{\rm i}0)).
\Eeqa
It is known that the integration by $x$ of (A.15) becomes zero~\cite{A16}.
This term transfers spectral intensity in the high energy region to the 
low energy region.
The expressions (A.12), (A.13) and (A.15) have convenient form
to calculate the 
particle excitation part, i. e., $x > 0$.
The expression can be transformed into a  suitable form to 
calculate the hole excitation case ($ x <0$ ) 
by changing the integration variable from  $y$
to $y+x$~\cite{C10}.


\begin{thebibliography}{9}
\bibitem{A14}
The QMC has limitation in application to the very strong interaction 
cases,
and the NRG is hard to apply to cases in which the 
the degeneracy factor of the impurity orbital,
$N_{f}$, is  large.
\bibitem{A57}
The added hybridization is given by the linear broken line:
it starts at 0.1 Ry and has peak at 0.3 Ry  with $5 \times 10^{-4}$ Ry 
and goes zero at 0.4 Ry for $\gamma$ Ce.
It is shifted to high energy side by 0.1 Ry in $\alpha$ Ce, 
and by -0.1 Ry in CeSb.
\bibitem{A58}
The change of trial HI from the preceding iteration 
step is multiplied by a factor 
$\frac{\Gamma^{2}}{(\VEP-E_{\rm F})^{2}+ \Gamma^{2}}$ with  
$\Gamma =0.04$ Ry.
\bibitem{A59}
When we compare the spectra to check the convergence,
averaging of spectra by neighboring 40 points are carried 
out.
Because we have used logarithmic mesh near the Fermi energy,
the width of the energy  mesh is about $10^{-7}$ Ry very 
near the Fermi energy.
The averaging process almost does not change the spectra very near
the Fermi energy, but change of relative intensity about 0.1 
is caused in higher energy region, 
$\mid \VEP-E_{\rm F}\mid \sim 0.3$ Ry.
\bibitem{D5}
When the Ce $5d$ level was shifted to high energy side to 
reduced the band overlapping, 
$(j=5/2)\Gamma_{7}$ became relatively stable.
\end{thebibliography}
\end{document}